\documentclass[aps,prb,twocolumn,showpacs,letterpaper,tighten,float,superscriptaddress,nofootinbib,reprint]{revtex4-2}
\usepackage{amssymb}
\usepackage{amsbsy}
\usepackage{amsmath}
\usepackage{mathrsfs}
\usepackage{epsfig}
\usepackage{graphicx}
\usepackage{array}
\usepackage{textcomp}
\usepackage{color}
\usepackage{braket}
\usepackage{hhline}
\usepackage{comment}
\usepackage{bm,hyperref}
\usepackage[justification=raggedright,font=small]{caption}
\usepackage[titletoc,toc,title]{appendix}
\usepackage[normalem]{ulem}
\usepackage[labelformat=simple]{subcaption}
\usepackage{tikz-cd}

\definecolor{myblue}{rgb}{.93, .93, 1}

\definecolor{darkgreen}{rgb}{0,0.7,0}

\newcommand{\beq}{\begin{equation}\begin{aligned}}
\newcommand{\eeq}{\end{aligned}\end{equation}}

\newcommand{\bpm}{\begin{pmatrix}}
\newcommand{\epm}{\end{pmatrix}}
\newcommand{\bmm}{\begin{matrix}}
\newcommand{\emm}{\end{matrix}}

\newcommand{\mbZ}{\mathbb{Z}}

\newcommand{\ii}{\mathbf{i}}

\graphicspath{{./figures/}}

\definecolor{NT}{rgb}{0.8,0,0.8}

\newcommand{\nocontentsline}[3]{}
\let\origcontentsline\addcontentsline
\newcommand\stoptoc{\let\addcontentsline\nocontentsline}
\newcommand\resumetoc{\let\addcontentsline\origcontentsline}

\begin{document}

\author{Pranay Gorantla}
\affiliation{Kadanoff Center for Theoretical Physics \& Enrico Fermi Institute, University of Chicago, Chicago, IL 60637, USA}
\affiliation{Department of Physics, Princeton University, Princeton, NJ 08544, USA}
\author{Abhinav Prem}
\affiliation{School of Natural Sciences, Institute for Advanced Study, Princeton, NJ 08540, USA}
\affiliation{Physics Program, Bard College, 30 Campus Road, Annandale-On-Hudson, NY 12504, USA}
\author{Nathanan Tantivasadakarn}
\affiliation{Walter Burke Institute for Theoretical Physics and Department of Physics, California Institute of Technology, Pasadena, CA, 91125, USA}
\author{Dominic J. Williamson}
\affiliation{School of Physics, The University of Sydney, NSW 2006, Australia}

\date{\today}

\title{String-Membrane-Nets from Higher-Form Gauging: An Alternate Route to $p$-String Condensation}

\date{\today}

\begin{abstract}
We present a new perspective on the $p$-string condensation procedure for constructing 3+1D fracton phases by implementing this process via the gauging of higher-form symmetries. Specifically, we show that gauging a 1-form symmetry in 3+1D that is generated by Abelian anyons in isotropic stacks of 2+1D topological orders naturally results in a 3+1D $p$-string condensed phase, providing a controlled non-perturbative construction that realizes fracton orders. This approach clarifies the symmetry principles underlying $p$-string condensation and generalizes the familiar connection between anyon condensation and one-form gauging in two spatial dimensions. We demonstrate this correspondence explicitly in both field theories and lattice models: in field theory, we derive the foliated field theory description of the $\mathbb{Z}_N$ X-Cube model by gauging a higher-form symmetry in stacks of 2+1D $\mathbb{Z}_N$ gauge theories; on the lattice, we show how gauging a diagonal 1-form symmetry in isotropic stacks of $G$-graded string-net models leads to string-membrane-nets hosting restricted mobility excitations. This perspective naturally generalizes to spatial dimensions $d \geq 2$ and provides a step towards building an algebraic theory of $p$-string condensation.
\end{abstract}

\maketitle

\tableofcontents


\section{Introduction}
\label{sec:intro}

A central motif in quantum many-body physics is the classification of all possible phases of matter in strongly correlated quantum systems. An important subset are gapped topological phases i.e., phases with an energy gap separating the ground state manifold from excited states (in the thermodynamic limit) and where local operators act trivially within the ground state manifold. The universal low-energy physics of such phases is expected to be captured by topological quantum field theory (TQFT), which provides a general framework for describing the topological braiding and exchange statistics between fractionalized excitations in arbitrary dimensions~\cite{wenreview}. In two spatial dimensions (2+1D), with only point-like localized excitations, the TQFT framework is generally believed to solve the classification problem in terms of modular tensor categories~\cite{Moore1989,kitaev}; in 3+1D, where additional extended loop-like excitations exist, there has likewise been progress in classifying those phases that permit a TQFT description, largely facilitated by exactly solvable models~\cite{walker2012,wan2015twisted,williamson2017ham,wen2018prx,wen2019prb,theo2020,theo2022class}.

The theoretical discovery of 3+1D exactly solvable models with fracton order~\cite{chamon,haah,castelnovo,yoshida,fracton1,fracton2,williamson}, characterized by topological excitations with restricted mobility (absent any disorder), has called for a re-examination of the TQFT paradigm. While gapped quantum phases in 3+1D with fracton order have much in common with their topologically ordered counterparts---including a nontrivial ground state degeneracy (GSD) on nontrivial manifolds, long-range entanglement, and point-like excitations with anyonic statistics---they evade the conventional TQFT description due to their sensitivity to the ambient geometry, such as the dependence of the GSD on the system size (see Refs.~\cite{fractonreview,fractonreview2} for a review), and hence display UV/IR mixing~\cite{gorantla2021uvirmixing}. There are now several exactly solvable lattice models that feature restricted mobility excitations~\cite{twisted,cagenet,premgauging,bulmashgauging,Tantivasadakarnsearching20,FractonCSBF20,JWfracton20,Shirley20,hybrid2021,TJV2,HsinSlagle21,defectnetworks,gorantla2023graphs}, with \emph{fracton} excitations referring to those which are fully immobile in isolation and \emph{lineons}/\emph{planons} referring to those which are mobile only along sub-dimensional lines/planes of the 3D manifold. Gapped fracton phases are an intense area of research across various fields, given their potential applications in quantum information storage and processing~\cite{haah2,Brown2019,albertespec,Song22,Davydova23,Zhang2022xcube}, anomalous quantum dynamics~\cite{kimhaah,prem}, interesting phase transitions~\cite{domhigherform,poon2021phase,you2022critical,Zhu23}, and unconventional quantum field theory descriptions~\cite{SlagleXcubeQFT,SlagleSMN,Slagle21,SeibergSymmetry,Seiberg:2020bhn,seiberg2021zn,gorantla2021villain}.

A remarkable property of phases with fracton order is that the restricted mobility of excitations is not imposed energetically, as is the case in kinetically constrained models, but rather results from unconventional emergent symmetries. For instance, one mechanism for engineering restricted mobility excitations is by gauging ``subsystem symmetries" i.e., symmetries that only act on subdimensional manifolds~\cite{fracton2,williamson,yizhi1,strongsspt,spurious,shirleygauging,Stephen2020,Shirley23}. A distinct approach for producing restricted mobility quasiparticles, dubbed ``$p$-string condensation," was introduced in Refs.~\cite{han,sagar} for the X-Cube model and later generalized in Ref.~\cite{cagenet} to realise a large class of ``cage-net" fracton models (see also Refs.~\cite{designer,SullivanPlanarpstring}). In this construction, layers of 2+1D topological orders are coupled together by condensing extended 1+1D strings (dubbed $p$-strings) composed of point-like excitations, and the braiding of other excitations with the $p$-strings in the un-condensed phase determines their mobility in the $p$-string condensed phase. A virtue of this construction, which generalizes the familiar notion of anyon condensation (see Ref.~\cite{burnellreview} for a review), is that it allows one to transparently read-off properties of the fracton phase given the underlying TQFT data of the 2+1D layers, and naturally leads to the defect network construction for fracton phases~\cite{defectnetworks,Song2023} as well as their description in terms of string-membrane-nets~\cite{SlagleSMN}.

In this paper, we formulate a different perspective on $p$-string condensation which illuminates the symmetry principles underlying this mechanism; namely, we show that gauging a higher-form symmetry of isotropic layers of 2+1D topologically ordered states naturally results in a 3+1D $p$-string condensed phase\footnote{Here, the distinction we make between gauging and condensation is that in the former, the 1-form symmetry is replaced by a  dual 1-form symmetry after gauging, whereas in the latter, the same symmetry remains, and the phase goes through a phase transition to the condensed phase.}. This procedure is in analogy with how gauging a (non-anomalous) 1-form symmetry in a 2+1D TQFT, generated by Wilson loops of some Abelian boson, results in a phase where that boson is condensed (i.e., identified with the vacuum super-selection sector). A similar procedure, when carried out in one lower dimensions (i.e., in 2+1D)--whereby a diagonal 0-form symmetry in stacks of 1+1D gapped theories is gauged--produces symmetry protected fracton phases (which are only robust to symmetric local perturbations), which we also briefly discuss. On the lattice, we consider stacks of 2+1D string-net models~\cite{Levin2005} and show that gauging a 3+1D 1-form symmetry generated by Wilson-lines of Abelian anyons from the underlying 2+1D TQFTs leads precisely to a $p$-string condensed theory of fractons. We also show how these principles can straightforwardly be applied in field theory and obtain the foliated field theory description of the 3+1D $\mbZ_N$ X-Cube model by gauging a higher-form symmetry generated by stacks of 2+1D $\mbZ_N$ gauge theories. Unlike the original formulation of $p$-string condensation, where a quantum phase transition separates the weakly coupled stacks of 2+1D TQFTs from the strongly coupled $p$-string condensed phase, this gauging procedure provides an exact gap-preserving map between these phases and is, as such, closer in spirit to the algebraic formulation of anyon condensation. Moreover, this gauging perspective illuminates various features and hidden dualities which intertwine topological and fracton orders, as well as their underlying symmetries in a nontrivial manner. We discuss these perspectives at length in a companion paper~\cite{forthcoming}: for instance, the ground state degeneracy of the X-Cube model can be simply understood as that of stacks of 2+1D Toric Codes from which (heuristically speaking) a 3+1D Toric Code has been eliminated via the higher-form gauging process.  We note that a distinct higher-form perspective on the X-Cube model was previously introduced in Ref.~\cite{Qi21}, where the fracton phase was interpreted as one with a spontaneously broken ``foliated" 1-form symmetry. 

The rest of this paper is organized as follows: in Sec.~\ref{sec:general}, we establish the higher-form gauging perspective on $p$-string condensation in general. In Sec.~\ref{sec:field}, we explicitly illustrate this correspondence within field theory; first, we show how the 2+1D $\mathbb Z_N$ Plaquette Ising model~\cite{fracton2,johnston2017} results from gauging a diagonal 0-form symmetry supported by stacks of 1+1D $\mbZ_N$ gauge theories. Next, we demonstrate how gauging a diagonal 1-form symmetry hosted by isotropic layers of 2+1D $\mbZ_N$ gauge theories produces the 3+1D $\mbZ_N$ X-Cube model~\cite{slagle3,seiberg2021zn}. In Sec.~\ref{sec:Cagenet}, we illustrate these ideas on the lattice: starting with isotropic stacks of 2+1D $G$-graded string-net models, we show that gauging a diagonal 1-form symmetry $\widehat{G}$ results in the condensation of $\widehat{G}$-valued $p$-string excitations and produces a 3+1D fracton phase. We conclude in Sec.~\ref{sec:cncls} with a discussion of open questions and future directions.


\section{$p$-string condensation via higher-form gauging}
\label{sec:general}

In this Section, we lay out the general picture for constructing 3+1D fracton phases by gauging a 1-form symmetry generated by anyonic string operators on stacks of 2+1D topological quantum field theories. In particular, we show that this procedure exactly produces a fracton phase in which $p$-strings--composed of Abelian anyons supported on the 2+1D layers--are condensed (we refer the reader to Ref.~\cite{cagenet} for a review of $p$-string condensation). Throughout, we assume lattice translation symmetry i.e., we assume that each layer supports an identical topological order and that the layers are isotropically arranged. These assumptions are for convenience, and much of our analysis carries through to gauging 1-form symmetries on layers of 2+1D TQFT that are neither translation invariant nor isotropic. 

\begin{figure}
    \centering
    \includegraphics[width=0.45\textwidth]{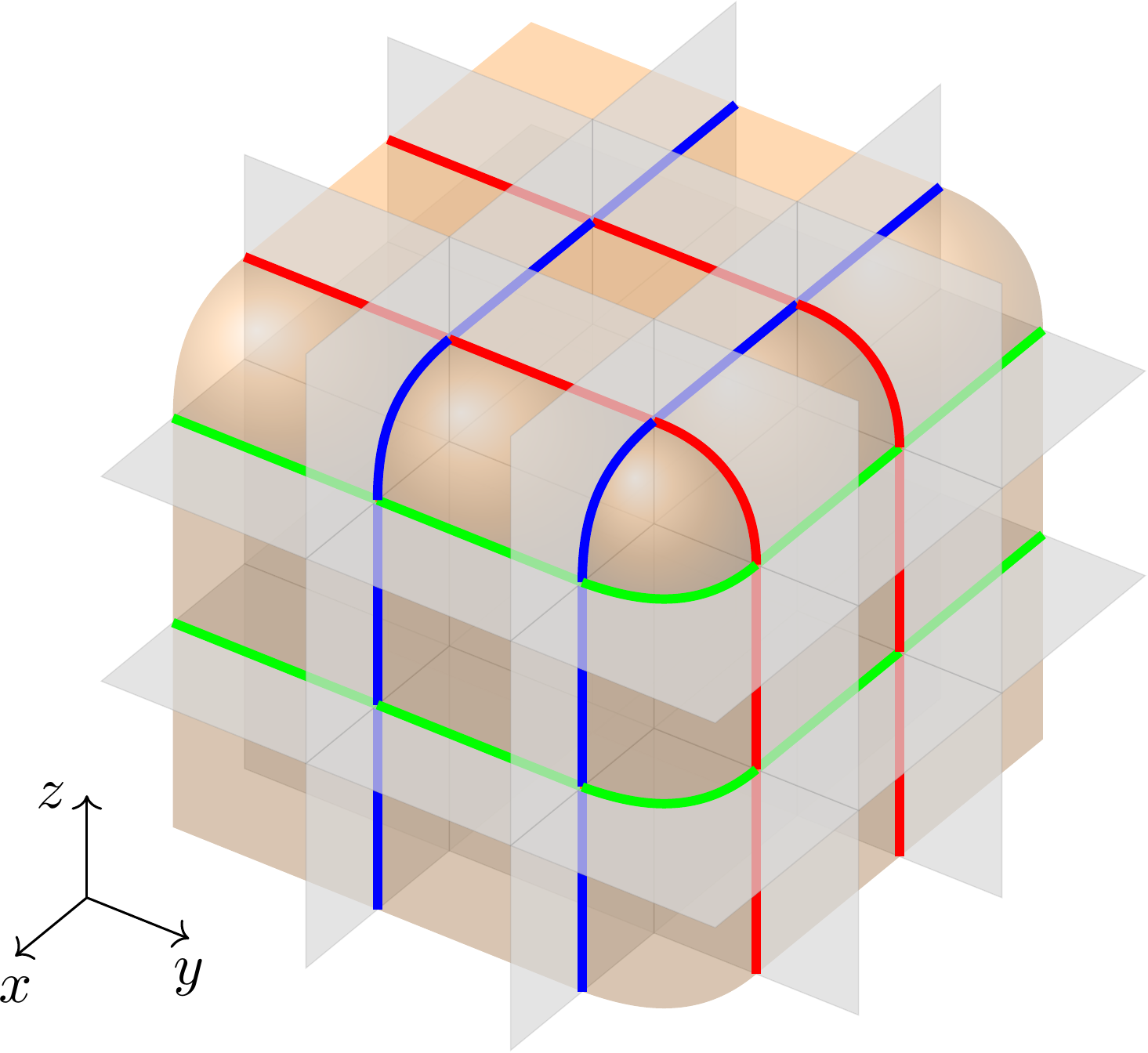}
    \caption{Each 2+1D layer $\ell$ (in grey) hosts an identical 2+1D topological order with some Abelian anyons $A$. The 1-form symmetry group $A_\ell^{(1)}$ in each layer is generated by string operators for $A$, illustrated here in green, blue, and red for layers along the $xy$, $xz$, and $yz$ planes respectively. The diagonal 1-form symmetry $A^{(1)}_{\mathrm{diag}}$ in the 3+1D bulk, which is generated by the layer 1-form symmetries, is represented by the closed orange surface.}
    \label{fig:mainfig}
\end{figure}

\paragraph*{Symmetries of a TQFT stack:} To describe fracton phases of matter that encompass cage-net models~\cite{cagenet}, we start from decoupled layers of 2+1D topological orders, which on the lattice are described by string-net models~\cite{Levin2005} (see Appendix~\ref{app:stringnet} for a brief overview). The symmetry group of a stack of string-net models corresponds to a codimension-1 foliated 1-form symmetry, following the terminology of Ref.~\cite{domhigherform}. Here, the symmetry is generated by string operators corresponding to Abelian anyons $A$ in the individual layers. We will use $A^{(1)}_\ell$ to denote this 1-form symmetry in a single layer $\ell$. Hence, the stack of layers inherits a formal codimension-1 foliated 1-form symmetry group structure $\prod_{\ell} A^{(1)}_\ell$, which contains a \textit{diagonal} subgroup $A^{(1)}_{\mathrm{diag}}$ that corresponds to a global 1-form symmetry in 3+1D~\cite{designer}. This subgroup is generated by symmetry operators that act on 2-cycles $z_{(2)}$ via a product of string operators, on the intersection of the 2-cycle with the layers $\ell$, taken from the relevant symmetry groups $A^{(1)}_\ell$ (see Fig.~\ref{fig:mainfig}). A 2-cycle is analogous to a membrane, and its intersection with each layer is analogous to a closed curve. Schematically, $U(z_{(2)}) := \prod_{\ell} U_\ell(z_{(2)}\cap \ell)$ for 1-form representations $U$ and $U_\ell$ in 3+1D and on the 2+1D layer $\ell$, respectively. 

The anyonic excitations of the stack are inherited from those of the layers. They can be organized into charge sectors under any subgroup $\widehat{G}$ of the 1-form symmetry $A^{(1)}_{\mathrm{diag}}$. We use a convention where the dual of an Abelian group $G$ is used to denote a subgroup of 1-form symmetries, while elements of the group $G$ denote line operators that are charged under the 1-form subgroup $\widehat{G}$. The charge sector $g\in G$ contains all product of anyons across the layers with charge labels $h_\ell$ that satisfy $\prod_\ell h_\ell = g$. On the other hand, a truncated diagonal symmetry operator applied to a 2-chain creates a composite $p$-string excitation, made up of $\chi\in \widehat{G}$ bosons on the intersection of the 2-chain's 1-cycle boundary with the layers $\ell$ (see Fig.~\ref{fig:p-string3d-trunc}). 

\paragraph*{$p$-string excitations and 1-form symmetry:} As mentioned above, $p$-string excitations are 1-dimensional composite excitations that are defined on 1-cycles in 3-dimensional space. Given a group of mutual $\widehat{G}$ bosons supported on the 2+1D layers $\ell$, and a $\widehat{G}$-valued 1-cycle $c_{(1)}$, the associated $p$-string excitation is a composite of $\widehat{G}$ bosons at the points of intersection between the 1-cycle and the layers $\ell$. Here, the boson that appears at each intersection point is determined by the $\widehat{G}$-label carried by the 1-cycle at that intersection point and the relative orientation of the 1-cycle and layer where they intersect. If the 1-cycle is labelled by $\chi \in \widehat{G}$  at the intersection point, and its orientation matches that of the layer, then a $\chi$ boson appears. If the orientations are opposing, then a $\chi^*$ boson appears. The above definition generalizes straightforwardly to allow different groups of bosons, isomorphic to $\widehat{G}$, and different topological orders in the layers. For simplicity, here we consider the same string-net model, with the same group of mutual bosons $\widehat{G}$, in each layer. 

Finally, note that a $p$-string excitation on a $\widehat{G}$-valued 1-cycle that is a 2-boundary can be created by applying a membrane operator on a $\widehat{G}$-valued 2-chain, whose boundary is the 1-cycle. 

\paragraph*{$p$-string condensed fracton theory:} Gauging the diagonal 1-form symmetry described above results in the condensation of the $\widehat{G}$-valued $p$-string excitations that are created at the boundary of truncated symmetry operators. The excitations of the emergent fracton order after $p$-string condensation can be classified in terms of the original anyons and the action of the $\widehat{G}$-valued 1-form symmetry as described below:
\begin{itemize}
    \item Closed $p$-string excitations, described by $\widehat{G}$-valued 1-cycles, are condensed and become equivalent to the vacuum superselection sector. Hence, any excitations that are related by fusion with a $p$-string become equivalent. The condensed $p$-strings correspond to extended objects that are charged under a dual $G$-valued 1-form symmetry.    
    \item Truncated $p$-string excitations, described by $\widehat{G}$-valued 1-chains $z_{(1)}\in Z_1(C,\widehat{G})$ (where $C$ denotes a cellulation) are condensed along their bulk but leave $\widehat{G}$-valued point-like excitations at their boundaries $\partial z_{(1)}$. 
    These are point-like gauge-fluxes of the gauged 1-form symmetry. 
    In isolation, these point-like excitations cannot pass through the string-net layers i.e., such a point-like excitation and its translate across a string-net layer are in distinct superselection sectors. 
    This results in the point-like excitations having restricted mobility. 
    For a $p$-string condensation on layers stacked in three orthogonal directions, these point-like excitations are fractons. 
    For layers stacked in two directions, they are lineons, and for a single stack, they are planons. 
    In all cases, a pair of point-like excitations that is created by $p$-string condensing a $\widehat{G}$ boson on a single layer is a planon with mobility in the same layer. 
    Fusion products of such planons with similarly oriented planes remain planons. 
    The fusion product of a pair of such planons on distinctly oriented planes is a lineon along the shared axis of the planes. 
    
    \item The $\mathcal{Z}(\mathcal{C})$ anyons in each string-net layer that have trivial 1-form charge, i.e.~those that braid trivially with the $\widehat{G}$-bosons in the layer, remain deconfined planons. 

    \item The $\mathcal{Z}(\mathcal{C})$ anyons in each string-net layer that have nontrivial 1-form charge, i.e.~braid nontrivially with the $\widehat{G}$-bosons in the layer, become confined. 
    The string operators for such anyons correspond to charges under a dual $G$-valued 1-form symmetry, which become extended gauge charges after $p$-string condensation. 
    These extended excitations are reducible, as a segment of such a string operator leads to a segment of the associated gauge charge excitation, in isolation. 
    Pairs of such anyons from different layers that have nontrivial individual 1-form charge but trivial total 1-form charge form partially mobile excitations. 
    If the layers intersect, they are lineon excitations along the common axis of the layers. 
    If the layers are parallel, they form planon excitations in the common plane of the layers. 
    More general bound-state clusters of anyons with nontrivial individual 1-form charge but trivial total 1-form charge are also possible, and their mobility depends on the orientation of the layers they are defined on. 
    In light of these bound states, we see that the extended gauge charge excitations can end on \textit{half-lineons}, which are confined particles. 
\end{itemize}
The gauged model has a dual $G$-valued 1-form symmetry and displays an interesting pattern of 1-form symmetry fractionalization on the cage-net fractons (see discussion in Sec.~\ref{sec:Cagenet}); we analyze this for the X-Cube model in our companion paper~\cite{forthcoming}, but leave the general case for future work.

While this Section has focused on $p$-string condensation via 1-form gauging in 3+1D, this procedure generalizes straightforwardly to any spatial dimension. In particular, a $d$-dimensional stack of $(d-1)$-dimensional gapped systems (where $d \geq 2$ is the spatial dimension), each with a $(d-2)$-form symmetry $G$, results in a $d$-dimensional theory with a codimension-1 foliated $(d-2)$-form symmetry $G_{\text{fol}}$ (in the lexicon of Ref.~\cite{domhigherform}). Gauging a diagonal $(d-2)$-form symmetry $G_\text{diag} \subset G_\text{fol}$, which is a global $(d-2)$-form symmetry in $d$-dimensions, realizes a phase that typically hosts excitations with restricted mobility.


\section{Field theory perspective}
\label{sec:field}


In this Section, we will illustrate the ideas laid out in Sec.~\ref{sec:general} through two concrete and familiar examples: the 2+1D $\mbZ_N$ Plaquette Ising model and the 3+1D $\mbZ_N$ X-Cube model. In both cases, $G = \mathbb Z_N$ is Abelian. Moreover, both models admit two complementary field theory descriptions: one is a foliated field theory~\cite{SlagleSMN,Slagle21,HsinSlagle21,Ohmori:2022rzz} that involves a standard set of fields in the bulk and on the layers, whereas the other is an exotic field theory~\cite{Seiberg:2020bhn,seiberg2021zn,gorantla2021villain} that uses higher-rank tensor gauge fields. We find it natural to demonstrate the ``$p$-string-condensation $\cong$ higher-form-gauging'' correspondence in the foliated field theory perspective, but also comment on the relation between the two perspectives. Throughout this Section, we work in the Euclidean signature with $\tau$ denoting Euclidean time and $x,y,z,\ldots$ denoting the spatial directions. We assume periodic boundary conditions in both space and time directions. We use $\mu,\nu,\ldots$ for spacetime indices and $i,j,\ldots$ for spatial indices. 

\subsection{A 2+1D example: $\mbZ_N$ plaquette Ising model}
\label{sec:2D}

Our first (warmup) example is the 2+1D $\mbZ_N$ plaquette Ising model, which we show how to produce by gauging a symmetry generated by stacks of 1+1D theories. Note that this procedure was previously carried out on the lattice in Ref.~\cite{domhigherform}. We begin with a stack of wires of 1+1D $\mathbb Z_N$ gauge theories along both the $x$ and $y$ directions embedded in 2+1D. The total Lagrangian is given by
\begin{align}
\label{2dstacklag}
    \mathcal L_\text{2Dstack} =& -\sum_{n_x=1}^{L_x} \frac{\ii N}{2\pi} \Phi^{(x)} dA^{(x)} \delta(x - n_x \varepsilon_x) dx
\\ \nonumber
&+ \sum_{n_y = 1}^{L_y} \frac{\ii N}{2\pi} \Phi^{(y)} dA^{(y)} \delta(y - n_y \varepsilon_y) dy~,
\end{align}
where $\varepsilon_i$ is the spacing between the wires orthogonal to the $i$-th spatial direction and $L_i$ is the number of such wires. Here, $\Phi^{(x)}=\Phi^{(x)}(\tau,y;n_x)$ and $A^{(x)}=A^{(x)}(\tau,y;n_x)$, i.e., $n_x$ labels a wire orthogonal to the $x$ direction and $y$ is the spatial coordinate along the wire. Similar comments apply to the fields on the wires orthogonal to the $y$ direction. One can interpret this as the low energy description of two perpendicular stacks of 1+1D $\mathbb Z_N$ Ising models.

\begin{figure}[t]
\centering
\includegraphics[width=0.4\textwidth]{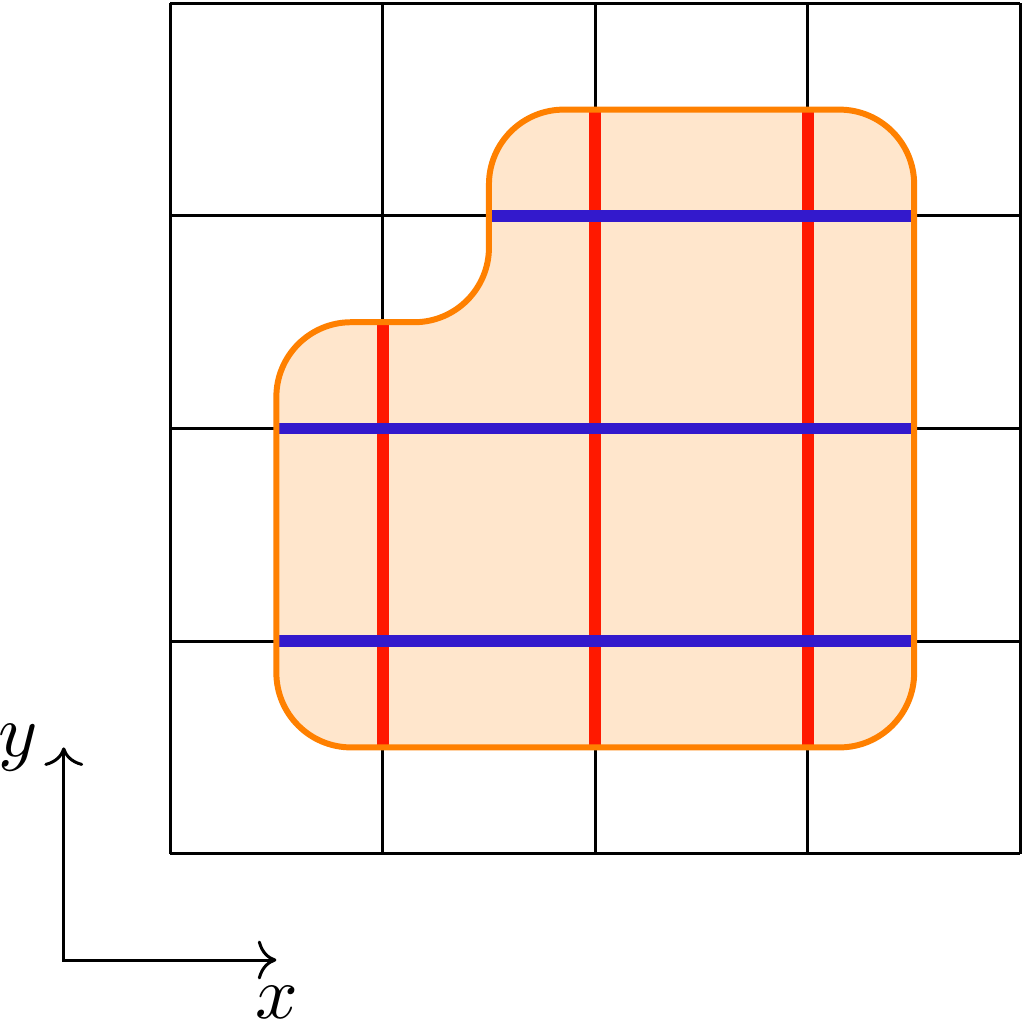}
\caption{The $p$-string operator Eq.~\eqref{eq:2d-pstring} in the 2+1D coupled-wire theory described by the Lagrangian Eq.~\eqref{2dpimlag1}: black lines represent the 1+1D wires in the 2+1D bulk. The orange, red, and blue curves are the Wilson line operators of $\hat a$, $A^{(x)}$, and $A^{(y)}$, respectively. When the orange curve is empty, i.e., when the orange surface spans the entire space, and the red and blue lines span all the wires, we get the diagonal 0-form symmetry operator Eq.~\eqref{eq:diag-0-form}.}
\label{fig:p-string2d}
\end{figure}

The theory specified by Eq.~\eqref{2dstacklag} has \emph{2-foliated} $\mathbb Z_N$ 0-form and 1-form global symmetries coming from the individual wires. They are generated by the Wilson operators of $A^{(i)}$ and the local operators $e^{\ii \Phi^{(i)}}$, respectively. We are interested in gauging the \emph{diagonal} $\mathbb Z_N$ 0-form symmetry generated by the operator
\beq\label{eq:diag-0-form}
\prod_{n_x=1}^{L_x} W^{(x)}(n_x) \prod_{n_y=1}^{L_y} W^{(y)}(n_y)~.
\eeq
This is a standard $\mathbb Z_N$ 0-form global symmetry in 2+1D and so, to gauge it, we couple the stack Lagrangian Eq.~\eqref{2dstacklag} to a dynamical 2+1D $\mathbb Z_N$ 1-form gauge field $a$:
\begin{align}
\label{2dpimlag1}
    \mathcal L_\text{2Dgauge} &= \sum_{n_x=1}^{L_x} \frac{\ii N}{2\pi} (d\Phi^{(x)} - a)A^{(x)} \delta(x - n_x \varepsilon_x) dx
\\ \nonumber
&- \sum_{n_y = 1}^{L_y} \frac{\ii N}{2\pi} (d\Phi^{(y)} - a) A^{(y)} \delta(y - n_y \varepsilon_y) dy \\ \nonumber
&+ \frac{\ii N}{2\pi} \hat a da~.
\end{align}
Here, $\hat a$ is a dynamical $U(1)$ gauge field which ensures that $a$ is a $\mathbb Z_N$ gauge field rather than a $U(1)$ gauge field. The gauge symmetry acts as
\begin{align}
    \Phi^{(i)} & \sim \Phi^{(i)} + 2\pi N^{(i)}(n_i) + \lambda~, \\ \nonumber
    A^{(i)} & \sim A^{(i)} + d\Lambda^{(i)}~, \\ \nonumber
    a & \sim a + d\lambda~, \\ \nonumber
    \hat a & \sim \hat a + d\hat \lambda + \sum_{n_x=1}^{L_x} \Lambda^{(x)} \delta(x - n_x \varepsilon_x) dx \\ \nonumber
    & - \sum_{n_y = 1}^{L_y} \Lambda^{(y)} \delta(y - n_y \varepsilon_y) dy~,
\end{align}
with $N^{(i)}(n_i) \in \mathbb Z$ an independent integer on each wire.

The Lagrangian Eq.~\eqref{2dpimlag1} provides the foliated field theory description of the spontaneously broken phase of the 2+1D $\mathbb Z_N$ plaquette Ising model. To see this, let us relate the operators of this theory to the operators of the $\mathbb Z_N$ plaquette Ising model (such foliated/exotic correspondences were explored before in Ref.~\cite{Ohmori:2022rzz}). Recall that on an $L_x \times L_y$ spatial lattice with periodic boundary conditions, the minimal (exotic) Lagrangian that captures this phase is (schematically) given by~\cite{Seiberg:2020bhn,gorantla2021villain}\footnote{\label{ftnt:disc-cont}Here, we find it convenient to treat the Euclidean time direction as continuous and the spatial directions as discrete. This is unlike the usual Euclidean setup where the time and spatial directions are either both continuous or both discrete. The discreteness in the spatial direction has its origin in the nonzero separations $\varepsilon_i$'s between the layers of the stack.}
\beq
\label{2dpimlag}
\mathcal L_\text{PIM} = \frac{\ii N}{2\pi} \phi^{xy} (\partial_\tau A_{xy} - \Delta_x \Delta_y A_\tau)~.
\eeq
This is the Lagrangian of the 2+1D $\mathbb Z_N$ hollow tensor gauge theory. The gauge symmetry is
\begin{align}
    \phi^{xy} & \sim \phi^{xy} + 2\pi w^{xy}_x(n_x) + 2\pi w^{xy}_y(n_y)~, \\ \nonumber
    A_\tau & \sim A_\tau + \partial_\tau \alpha~, \\ \nonumber
    A_{xy} & \sim A_{xy} + \Delta_x \Delta_y \alpha~,
\end{align}
where $w^{xy}_i(n_i) \in \mathbb Z$. While $\phi^{xy},A_{xy}$ are placed on the sites labelled by $(n_x,n_y)$, $A_\tau$ is placed on the plaquettes labelled by $(n_x+\tfrac12,n_y+\tfrac12)$. The defect of $A_\tau$ is a fracton, and a dipole of fractons separated along the $x$ ($y$) direction can be moved in the $y$ ($x$) direction using the membrane operator of $A_{xy}$. Stretching the membrane operator in the $x$ or $y$ direction gives the generators of the magnetic $\mathbb Z_N$ subsystem symmetry. On the other hand, the electric $\mathbb Z_N$ subsystem symmetry is generated by $e^{\ii \phi^{xy}}$. The correspondence between the operators and defects of the two presentations Eq.~\eqref{2dpimlag1} and Eq.~\eqref{2dpimlag} is~\cite{Ohmori:2022rzz}:
\begin{widetext}
    \begin{align}
        \text{Local Operator:}\quad & \exp\left[\ii \phi^{xy}(\tau,n_x,n_y)\right] \longleftrightarrow \exp\left[\ii \Phi^{(x)}(\tau,n_y\varepsilon_y;n_x) - \ii \Phi^{(y)}(\tau,n_x\varepsilon_x;n_y)\right]~,
    \end{align}
    \begin{align}
    \text{Fracton Defect:}\quad & \exp\left[\ii \oint d\tau~ A_\tau(\tau,n_x+\tfrac12,n_y+\tfrac12)\right] \\ \nonumber
    \quad & \longleftrightarrow \exp\left[\ii \oint d\tau~ \hat a_\tau(\tau,x,y)\right]~,\quad \text{where}\quad n_i = \lfloor x_i/\varepsilon_i \rfloor~,
    \end{align}
    \begin{align}
    \label{pimcorr1}
    \text{Membrane Operator:}\quad& \exp\left[\ii \sum_{n_x = p_x}^{q_x-1} \sum_{n_y = p_y}^{q_y-1} A_{xy}(\tau,n_x,n_y)\right] \\ \nonumber
\quad & \longleftrightarrow \exp\left[\ii \int_{x_1}^{x_2} dx~ [\hat a_x(\tau,x,y_2) - \hat a_x(\tau,x,y_1)] - \ii \sum_{n_x=p_x}^{q_x-1} \int_{y_1}^{y_2} dy~A^{(x)}_y(\tau,y;n_x)\right] \\ \nonumber
\qquad & = \exp\left[\ii \int_{y_1}^{y_2} dy~ [\hat a_y(\tau,x_2,y) - \hat a_x(\tau,x_1,y)] + \ii \sum_{n_y=p_y}^{q_y-1} \int_{x_1}^{x_2} dx~A^{(y)}_x(\tau,x;n_y)\right]~,
    \end{align}
where $p_i = \lceil x_{i,1}/\varepsilon_i \rceil$ and $q_i =  \lceil x_{i,2}/\varepsilon_i \rceil$. In fact, the equality of the last two lines of Eq.~\eqref{pimcorr1} is precisely equivalent to $p$-string condensation, i.e, the $p$-string operator acts trivially on the ground states:
\beq\label{eq:2d-pstring}
\exp\left[ \ii \oint_\Gamma \hat a + \ii \sum_{n_x=1}^{L_x} \int_{\Gamma_x(n_x)} A^{(x)}(n_x) + \ii \sum_{n_y=1}^{L_y} \int_{\Gamma_y(n_y)} A^{(y)}(n_y)\right] = 1~,
\eeq
\end{widetext}
where $\Gamma$ is a curve in the $xy$ plane, and $\Gamma_i(n_i)$ is the portion of the $n_i$-th wire orthogonal to $i$-th spatial direction contained within $\Gamma$. See Fig.~\ref{fig:p-string2d} for an illustration of the $p$-string operator.


\subsection{3+1D X-Cube field theory}
\label{sec:Xcube}

Our second (main) example is the 3+1D $\mbZ_N$ X-Cube model. We begin with an isotropic stack of 2+1D $\mathbb Z_N$ gauge theories in 3+1D spaced equally along the $x$, $y$, and $z$ directions. The total Lagrangian is given by
\beq
\label{3dstacklag}
\mathcal L_\text{3Dstack} &= \sum_i\sum_{n_i=1}^{L_i} \frac{\ii N}{2\pi} \hat A^{(i)} dA^{(i)} \delta(x_i - n_i \varepsilon_i) dx_i~,
\eeq
where $\varepsilon_i$ is the spacing between the layers orthogonal to the $i$-th spatial direction and $L_i$ is the number of such layers. Here, $\hat A^{(x)}=\hat A^{(x)}(\tau,y,z;n_x)$ and $A^{(x)}=A^{(x)}(\tau,y,z;n_x)$, i.e., $n_x$ labels a layer orthogonal to the $x$ direction, and $y,z$ are the spatial coordinates along the layer. Similar comments apply to the fields on the layers orthogonal to the $y$ and $z$ directions. One can interpret this as the low energy description of three perpendicular stacks of 2+1D $\mbZ_N$ Toric Codes.

The theory Eq.~\eqref{3dstacklag} has two \emph{3-foliated} $\mathbb Z_N$ 1-form global symmetries---electric and magnetic---coming from the individual layers. They are generated by the Wilson lines of $\hat A^{(i)}$ and $A^{(i)}$ respectively. We are interested in gauging the \emph{diagonal} electric 1-form symmetry generated by operators of the form
\begin{align}
\label{3ddiagWil}
\hat{\mathcal W}(\Sigma) &= \prod_i \prod_{n_i=1}^{L_i} \hat W^{(i)}[\Gamma_i(n_i)]
\\ \nonumber
&= \exp\left[ \ii \sum_i \sum_{n_i=1}^{L_i} \int_{\Gamma_i(n_i)} \hat A^{(i)}(n_i)\right]~,
\end{align}
where $\Sigma$ is a closed surface and $\Gamma_i(n_i)$ is the intersection of $\Sigma$ with the $n_i$-th layer orthogonal to the $i$-th direction. This is a standard $\mathbb Z_N$ 1-form global symmetry in 3+1D and so, to gauge it, we couple the stack Lagrangian Eq.~\eqref{3dstacklag} to a dynamical 3+1D $\mathbb Z_N$ 2-form gauge field $b$:
\begin{align}
\label{3dxclag1}
\mathcal L_\text{3Dgauge} &= \sum_i\sum_{n_i=1}^{L_i} \frac{\ii N}{2\pi} \hat A^{(i)} (dA^{(i)}-b) \delta(x_i - n_i \varepsilon_i) dx_i \\ \nonumber
&+ \frac{\ii N}{2\pi} b da~.
\end{align}
Here, $a$ is a dynamical $U(1)$ gauge field which ensures that $b$ is a $\mathbb Z_N$ 2-form gauge field rather than a $U(1)$ 2-form gauge field. The gauge symmetry acts as
\begin{align}
A^{(i)} & \sim A^{(i)} + d\Lambda^{(i)} + \rho~, \\ \nonumber
\hat A^{(i)} & \sim \hat A^{(i)} + d \hat \Lambda^{(i)}~, \\ \nonumber
b & \sim b + d\rho~, \\ \nonumber
a & \sim a + d \lambda + \sum_i\sum_{n_i=1}^{L_i} \hat \Lambda^{(i)} \delta(x_i - n_i \varepsilon_i) dx_i~.
\end{align}
Note that the equation of motion of $b$ makes the operator in Eq.~\eqref{3ddiagWil} trivial, which is consistent with the fact that we gauged the symmetry it generates.

The Lagrangian Eq.~\eqref{3dxclag1} is the foliated field theory that captures the low energy phase of the 3+1D $\mathbb Z_N$ X-Cube model~\cite{Slagle21,SlagleSMN,HsinSlagle21}. Let us relate it to the higher-rank tensor gauge theory (i.e., exotic field theory) description of the same phase~\cite{Ohmori:2022rzz}. Recall that on an $L_x \times L_y \times L_z$ spatial lattice with periodic boundary conditions, the minimal (exotic) Lagrangian that captures this phase is (schematically) given by~\cite{seiberg2021zn,gorantla2021villain}(see Footnote~\ref{ftnt:disc-cont})
\begin{align}
\label{3dxclag}
\mathcal L_\text{XC} &= \frac{\ii N}{2\pi} \sum_{\text{cyclic} \atop i,j,k} A_{ij} (\partial_\tau \hat A^{ij} - \Delta_k \hat A^{k(ij)}_\tau) \\ \nonumber
&+ \frac{\ii N}{2\pi} A_\tau \sum_{i<j} \Delta_i \Delta_j \hat A^{ij}~.
\end{align}
This is the Lagrangian of the 3+1D $\mathbb Z_N$ hollow tensor gauge theory. The gauge symmetry is
\begin{align}
A_\tau & \sim A_\tau + \partial_\tau \alpha~, \\ \nonumber
A_{ij} & \sim A_{ij} + \Delta_i \Delta_j \alpha~, \\ \nonumber
\hat A^{i(jk)}_\tau & \sim \hat A^{i(jk)}_\tau + \partial_\tau \hat \alpha^{i(jk)}~, \\ \nonumber
\hat A^{ij} & \sim \hat A^{ij} + \Delta_k \hat \alpha^{k(ij)}~.
\end{align}
Here, $A_{ij},\hat A^{ij}$ are placed on $k$-links, $A_\tau$ is placed on cubes, and $\hat A^{k(ij)}_\tau$ is placed on sites. The defect of $A_\tau$ is a fracton, and a dipole of fractons separated along the $x$ direction can be moved in the $yz$ plane using the membrane operator of $A_{ij}$. The defect of $\hat A^{z(xy)}_\tau$ is a $z$-lineon, and a dipole of $z$-lineons separated along the $x$ direction can be moved in the $yz$ plane using the line operators of $\hat A^{ij}$. The membrane operators of $A_{ij}$ generate the magnetic $\mathbb Z_N$ subsystem symmetry, where the line operators of $\hat A^{ij}$ generate the electric $\mathbb Z_N$ subsystem symmetry.

\begin{figure}[t]
\centering
\includegraphics[scale=0.3]{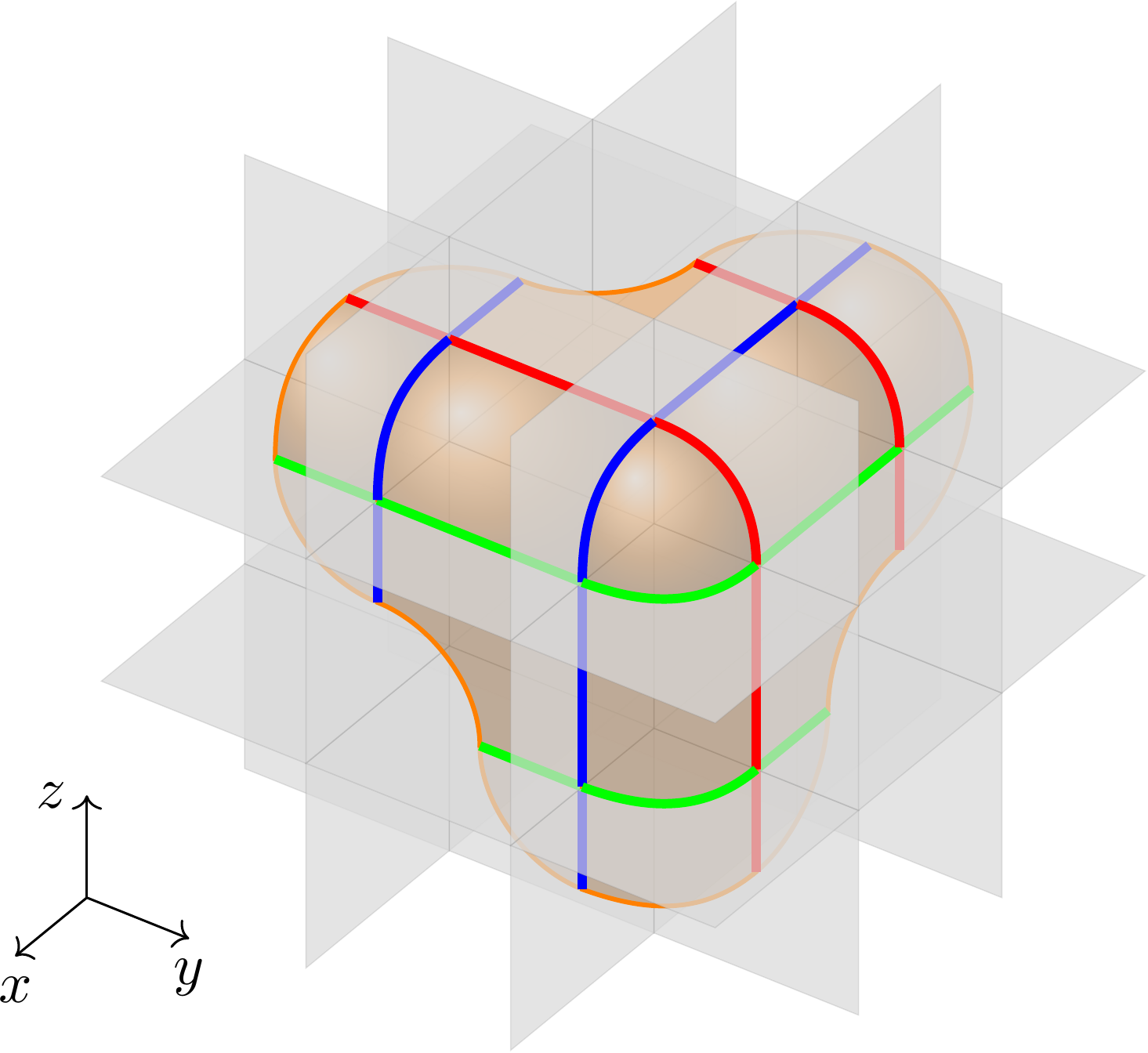}
\caption{The $p$-string operator Eq.~\eqref{eq:3d-pstring} in the 3+1D coupled-layer theory described by the Lagrangian Eq.~\eqref{3dxclag1}: the gray planes represent the 2+1D layers in the 3+1D bulk. The red, blue, and green curves are the Wilson line operators of the layer gauge fields $\hat A^{(x)}$, $\hat A^{(y)}$, and $\hat A^{(z)}$, respectively, whereas the orange line is the Wilson line operator of the bulk gauge field $a$. The orange surface is an artefact---for any given orange curve (in the trivial homology class), the choice of the orange surface is arbitrary, and this choice determines the red, blue, and green curves. When the orange curve is empty, i.e., when the orange surface is closed, we get the diagonal electric 1-form symmetry operator $\hat{\mathcal W}(\Sigma)$ in Eq.~\eqref{3ddiagWil}.}
\label{fig:p-string3d-trunc}
\end{figure}

The correspondence between the operators and defects of the two presentations Eq.~\eqref{3dxclag} and Eq.~\eqref{3dxclag1} is:
\begin{widetext}
\begin{align}
\text{Fracton Defect:}\quad& \exp\left[\ii \oint d\tau~ A_\tau(\tau,n_x+\tfrac12,n_y+\tfrac12,n_z+\tfrac12)\right] \\ \nonumber
\longleftrightarrow \, & \exp\left[\ii \oint d\tau~ a_\tau(\tau,x,y,z)\right]~,\quad \text{where}\quad n_i = \lfloor x_i/\varepsilon_i \rfloor~,
\end{align}
\begin{align}
\text{$z$-lineon Defect:}\quad& \exp\left[\ii \oint d\tau~ \hat A^{z(xy)}_\tau(\tau,n_x,n_y,n_z)\right] \\ \nonumber
\longleftrightarrow \, & \exp\left[\ii \oint d\tau~[A_\tau^{(x)}(\tau,n_y\varepsilon_y,n_z \varepsilon_z;n_x) - A_\tau^{(y)}(\tau,n_x\varepsilon_x,n_z \varepsilon_z;n_y)]\right]~,
\end{align}
\begin{align}
\label{xccorr1}
\text{Membrane Operator:}\quad& \exp\left[\ii \sum_{n_x = p_x}^{q_x-1} \sum_{n_y = p_y}^{q_y-1} A_{xy}(\tau,n_x,n_y,n_{z,0}+\tfrac12)\right] \\ \nonumber
\longleftrightarrow\,& \exp\left[\ii \int_{x_1}^{x_2} dx~ [a_x(\tau,x,y_2,z_0) - a_x(\tau,x,y_1,z_0)] - \ii \sum_{n_x=p_x}^{q_x-1} \int_{y_1}^{y_2} dy~\hat A^{(x)}_y(\tau,y,z_0;n_x)\right] \\ \nonumber
=\, & \exp\left[\ii \int_{y_1}^{y_2} dy~ [a_y(\tau,x_2,y,z_0) - a_x(\tau,x_1,y,z_0)] - \ii \sum_{n_y=p_y}^{q_y-1} \int_{x_1}^{x_2} dx~\hat A^{(y)}_x(\tau,x,z_0;n_y)\right]~, \\ \nonumber
&\qquad \text{where} \quad p_i = \lceil x_{i,1}/\varepsilon_i \rceil~,\quad q_i =  \lceil x_{i,2}/\varepsilon_i \rceil~, \quad \text{and} \quad n_{z,0} = \lfloor z_0/\varepsilon_z \rfloor~,
\end{align}
\begin{align}
\text{Line Operator:}\quad& \exp\left[\ii \sum_{n_z = p_z}^{q_z-1} \hat A^{xy}(\tau,n_x,n_y,n_z+\tfrac12)\right] \\ \nonumber
\longleftrightarrow \, & \exp\left[\ii \int_{p_z \varepsilon_z}^{q_z \varepsilon_z} dz~[A_z^{(x)}(\tau,n_y\varepsilon_y,z;n_x) - A_z^{(y)}(\tau,n_x\varepsilon_x,z;n_y)]\right]~.
\end{align}
\end{widetext}
In fact, the equality on the right hand side of the correspondence of the membrane operator in Eq.~\eqref{xccorr1} is equivalent to $p$-string condensation, i.e, the $p$-string operator acts trivially on the ground states:
\beq\label{eq:3d-pstring}
\exp\left[ \ii \oint_\Gamma a + \ii \sum_i \sum_{n_i=1}^{L_i} \int_{\Gamma_i(n_i)} \hat A^{(i)}(n_i)\right] = 1~,
\eeq
where $\Gamma$ is a closed spatial curve and $\Gamma_i(n_i)$ is an open spatial curve in the $n_i$-th layer orthogonal to the $i$-th spatial direction with endpoints on $\Gamma$. This follows from the equation of motion of $b$ in Eq.~\eqref{3dxclag1} (see Fig.~\ref{fig:p-string3d-trunc} for an illustration of the $p$-string operator). One can also simply write down the so-called ``belt" and ``cage" operators, which are composed of the above membrane and line operators, respectively. For example, the Wilson surface operator of the 2-form gauge field $b$ precisely corresponds to the cage-operator of the X-Cube model that detects an isolated fracton~\cite{Ohmori:2022rzz}.

\section{General cage-net construction}
\label{sec:Cagenet}

In this Section, we discuss the implementation of our 1-form symmetry gauging procedure to obtain a $p$-string condensed phase on the lattice as well as generalizations thereof. Specifically, we consider isotropic stacks of 2+1D $G$-graded string-net models and discuss how gauging a 1-form symmetry in 3+1D produces a fracton phase that is equivalent to a $p$-string condensed phase. For the sake of providing a self-contained discussion, string-net models are reviewed in Appendix~\ref{app:stringnet}.

\subsection{Cage-net models from 1-form gauging}

\paragraph*{1-form symmetry of a string-net stack:} The diagonal 1-form symmetry group of a stack of topological orders that admit gapped boundaries can be represented concretely using string-net models with a $G$-graded input category $\mathcal{C}_{G}$ on a set of layers $\ell$. 
For simplicity, we consider a cubic lattice cellulation of 3-space $C$, with layers $\ell$ corresponding to the $xy$, $yz$, and $xz$ lattice planes. 
It is simple to extend the construction to exclude lattice planes with a given orientation, or to include more general lattice planes.
We consider string-net models defined on square lattices, with resolved vertices, stacked onto the square sublattice layers $C_{\ell}$ of the cubic lattice cellulation.  
We describe the diagonal 1-form symmetry on the lattice model using 1-cocycles, as the $\widehat{G}$-boson string operators are naturally defined on the dual lattice of each string-net layer. 
Here, 1-cocycles are Poincar\'e dual to 2-cycles which describe membranes on the dual lattice. 
The diagonal 1-form $\widehat{G}$ symmetry has the following on-site representation
\begin{align}
    U:\quad Z^1(C,\widehat{G}) &\rightarrow \mathcal{U}(\mathcal{H}) \nonumber
    \\
    z^{(1)} &\mapsto U(z^{(1)}) := \prod_{e} U_e({z^{(1)}_e})
    ,
\end{align}
where $U_e(\chi)=\widehat{\chi}_{e,{\ell}} \widehat{\chi}_{e,{\ell'}}$ (see Eq.~\eqref{eq:chiop}) and where $\ell,\ell'$ denote the distinct layers that pass through edge $e$. 
Truncated symmetry operators are described by $\widehat{G}$-valued 1-cochains
\begin{align}
    U:\quad A^1(C,\widehat{G}) &\rightarrow \mathcal{U}(\mathcal{H}) \nonumber
    \\
    a^{(1)} &\mapsto U(a^{(1)}) := \prod_{e} U_e({a^{(1)}_e})
    ,
\end{align}
where the on-site representation $U_e$ is the same as above. 
The 1-cochain operator $U(a^{(1)})$ creates a $p$-string described by the coboundary $\delta a^{(1)}$, with $(\delta a^{(1)})_p$ bosons on the plaquettes $p\in C$. 
Plaquettes along the region where the symmetry operator is truncated support nontrivial $\widehat{G}$ bosons, while the remainder of the plaquettes host no excitations. 
The 2-cocycle $\delta a^{(1)}$ is Poincar\'e dual to a $\widehat{G}$-valued 1-cycle, matching the description of $p$-strings introduced in Sec.~\ref{sec:general} above. 

\paragraph*{Direct $p$-string condensation:} We briefly review the original $p$-string condensation procedure as introduced in Refs.~\cite{han,sagar} and generalized in Ref.~\cite{cagenet}. It is possible to drive a $p$-string condensation transition in a Hamiltonian for decoupled 2+1D layers of topological orders by adding a term to the Hamiltonian governing the decoupled stacks that proliferates the $p$-string excitations as its strength is increased. For $\widehat{G}$-valued $p$-strings, and layers given by topological orders that admit gappable boundaries, this can be achieved by using $G$-graded string-net models on the layers and coupling them by local $U_e(\chi)=\widehat{\chi}_{e,{\ell}} \widehat{\chi}_{e,{\ell'}}$ operators (see Appendix~\ref{app:stringnet}). These operators simultaneously create a pair of anyons each in two intersecting 2+1D string-net layers. The perturbed Hamiltonian is given by
\begin{align}
    {H}_\lambda = (1-\lambda) \sum_{\ell} H^{(\ell)}_{\mathrm{SN}} + \lambda V_{\widehat{G}} ,
\end{align}
where $H_{\mathrm{SN}}$ is the unperturbed string-net Hamiltonian for each layer $\ell$, $V_{\widehat{G}}=-\sum_e \frac{1}{|\widehat{G}|}\sum_{\chi \in \widehat{G}}U_e(\chi)$ couples the layers locally, and $\lambda\in [0,1]$. The $V_{\widehat{G}}$ coupling between strings from different layers energetically favors the trivial charge sector, under $U_e(\chi)$, on each edge. The above Hamiltonian undergoes a $\widehat{G}$-valued $p$-string condensation phase transition to a cage-net model as $\lambda$ increases (see Ref.~\cite{cagenet} for more details).

\paragraph*{$p$-string condensation via gauging 1-form symmetry:} The $p$-string condensation described above can alternatively be implemented by gauging the diagonal 1-form $\widehat{G}$ symmetry of a stack of $G$-graded string-net layers. As described in Sec.~\ref{sec:general}, this occurs because gauging the diagonal 1-form $\widehat{G}$ symmetry condenses the defects that are created by truncated symmetry operators. These defects are precisely $\widehat{G}$-valued $p$-strings.

To gauge the diagonal 1-form $\widehat{G}$ symmetry, we introduce a generalized gauge field $\mathbb{C}[\widehat{G}]$ degree of freedom onto each plaquette $p\in C$ in state $\ket{+}$. We measure the set of projection operators 
\begin{align}
    \label{eq:GGauss2}
    \Pi_e(g) := \frac{1}{|\widehat{G}|}\sum_{\chi \in \widehat{G}} \chi^*(g) U^\dagger_{p_1}(\chi) U^\dagger_{p_2}(\chi) U_e(\chi) U_{p_3}(\chi) U_{p_4}(\chi) ,
\end{align}
for $g\in G$, on every edge $e$, and for adjacent plaquettes $p_1,p_2,p_3,p_4$ in $C$. 
Here $p_1,p_2,$ are the plaquettes with orientation matching $e$ and $p_3,p_4,$ are the plaquettes with orientation opposite to $e$. 
The edge operators are defined above and the plaquette operators are similarly defined: $U_p(\chi) \ket{g} = \chi(g) \ket{g}$. 
The purpose of these measurements is to detect violations of the generalized Gauss's law for the 1-form symmetry. 
This follows from the action of the $\Pi_e(g)$ operators in the basis of $G$ group elements on plaquettes $p_1,p_2,p_3,p_4$, and $\mathcal{C}$ string types on the edge $e$,
\begin{align}
    &\Pi_e(f) \ket{(g_1)_{p_1}(g_2)_{p_2} (s_h t_k)_e (g_3)_{p_3} (g_4)_{p_4}} 
    \nonumber \\
    &= \delta_{f^{-1} g_1^{-1}g_2^{-1} h k g_3g_4}\ket{(g_1)_{p_1}(g_2)_{p_2} (s_h t_k)_e (g_3)_{p_3} (g_4)_{p_4}} ,
    \nonumber
\end{align}
where $f,g_1,g_2,g_3,g_4,h,k\in G$, $s_h\in \mathcal{C}_h,$ and $t_k\in\mathcal{C}_k$. 

The standard 1-form gauging procedure again corresponds to applying the $\Pi_e(1)$ projector to all edges. 
This can be achieved via post-selection or by applying a suitable Abelian byproduct operator after measurement. 
On the subspace of 1-form symmetric states, the measurement outcomes $g_e$ must form $G$-valued 1-cycles on $C$. 
Furthermore, for states that satisfy the nonlocal 1-form symmetry operators corresponding to nontrivial homology classes, the measurement outcomes must form a $G$-valued 1-boundary on $C$. 
Hence, there is a 2-chain $h_{(2)}$ on $C$ that satisfies $(\partial h_{(2)})_e = g_e$. 
This defines a byproduct operator $\prod_p L_p\big((h_{(2)})_p\big)$--see the discussion regarding gauging 1-form symmetries above. Again, note that any choice of 2-chain leads to the same outcome. The gauging of 1-form symmetric operators proceeds analogously to the gauging procedure outlined in Appendix~\ref{app:stringnet}. The only point of difference is that now, when gauging an operator, there are multiple choices of the 2-chain appearing in Eq.~\eqref{eq:GuugingM} that differ by local 2-boundaries. By convention, we choose a 2-chain with minimal support. 

\paragraph*{Cage-net model from gauging:} Gauging the $\widehat{G}$-valued 1-form symmetry of the stacked string-net layer Hamiltonian is performed term-wise on the vertex and plaquette operators, which are treated as sums of $B_p^g$ operators. 
This results in the gauged model
\begin{align}
    H_{\mathrm{SMN}}=\sum_{\ell} H_{\mathrm{GSN}}^{(\ell)} 
    - \Delta \sum_{e} \Pi_e(1)
    - \Delta' \sum_{c} F_c, 
\end{align}
where we have separated out the gauged string-net layers, Gauss's law projectors, defined in Eq.~\eqref{eq:GGauss2}, and terms that energetically enforce a zero-flux condition, respectively. 
All the local terms in the above Hamiltonian commute. 
The gauged string-net layer Hamiltonians are similar to those presented in Eqs.~\eqref{eq:GSN}, and~\eqref{eq:GSNH}, except we do not include the Gauss's law projector terms
\begin{align}
    H_{\mathrm{GSN}}^{(\ell)} &= - \sum_{v\in \ell} \mathcal{G}(A_v^{(1)})+\mathcal{G}(A_v^{(2)}) - \sum_{p \in \ell} \frac{1}{|G|}\sum_{g\in G}  \mathcal{G}(B_p^g) 
    \nonumber 
    \\
    &= - \sum_{v \in \ell} A_v^{(1)}+A_v^{(2)} - \sum_{p \in \ell} \frac{1}{|G|} \sum_{g \in G}   L_p(g) B_p^g ,
\end{align}
where the vertex terms are understood to act on edge degrees of freedom associated to layer $\ell$. 
The $F_c$ terms that energetically penalize generalized gauge-fluxes are defined on cubes $c$
\begin{align}
    \label{eq:SMNCube}
    F_c = \frac{1}{|G|} \sum_{g\in G}\prod_{p\in c} L_p^{\pm 1}(g) ,
\end{align}
where the $(\pm 1)$ superscript is determined by the relative orientation of $p$. 
It is $(+1)$ when the orientation of $p$ matches $c$, and $(-1)$ otherwise. 
Each operator appearing in the sum above corresponds to a 2-boundary of a 3-chain with a single non-identity group element on the cube $c$. 
The $F_c$ terms energetically enforce a zero-gauge-flux condition. 
Within the zero-gauge-flux subspace, the choice of 2-chain when gauging local operators does not matter in the following sense: any operators obtained by gauging the same initial operator with different choices of 2-chains act identically within the zero-gauge-flux subspace. 

The gauged model is a string-membrane-net model, corresponding to a stack of string-net layers that have undergone $\widehat{G}$-valued $p$-string condensation via coupling to $\mathbb{C}[G]$ membrane degrees of freedom living on faces. 
For string-net models that correspond to layers of $\mathbb{Z}_N$ Toric Codes, our construction reproduces the string-membrane-net models introduced in Ref.~\cite{SlagleSMN}. 
The gauged model has a dual $G$-valued 1-form symmetry
\begin{align}
    L:\quad Z_2(C,G) &\rightarrow \mathcal{U}(\mathcal{H}) \nonumber
    \\
    z_{(2)} &\mapsto L(z_{(2)}) := \prod_{p} L_p\big({(z_{(2)})_p}\big)
    .
\end{align}
The defects obtained by truncating this symmetry correspond to generalized gauge charges, which are 1-dimensional extended objects described by $G$-valued 1-cycles. They are generated by gauging string operators from nontrivial $G$-sectors in the string-net layers. We show below, via a gLU mapping to the cage-net model, that these string excitations are decomposable composite excitations. Each string can be interpreted as half of a lineon string operator (see discussion below). For the particular case of the X-Cube model, we discuss this interesting form of 1-form symmetry fractionalization in our companion paper~\cite{forthcoming}, but leave a deeper investigation of the 1-form symmetry enriched structure of such string-membrane-net models to future work.

The emergent fracton order that results from the $p$-string condensation induced by gauging the string-net layers follows the general discussion presented in Sec.~\ref{sec:general}. Moreover, the string-membrane-net models we have defined are phase equivalent to cage-net models~\cite{cagenet} under generalized-local-unitary (gLU)~\cite{Chen2010} circuits. The gLU circuits that map the cage-net model to the string-membrane-net model are 
\begin{align}
\label{eq:setent}
    U_{\mathrm{3DSET}}=\prod_{\ell} U_{\mathrm{SET}}^{(\ell)}
\end{align}
where the gLU operator $U_{\mathrm{SET}}^{(\ell)}$ defined in Eq.~\eqref{eq:USET} acts on the plaquettes and edges within each layer $\ell$. 
Conjugating the string-membrane-net Hamiltonian by the SET entangler Eq.~\eqref{eq:setent} yields the cage-net Hamiltonian 
\begin{widetext}
    \begin{align}
    U_{\mathrm{3DSET}}^\dagger H_{\mathrm{SMN}} U_{\mathrm{3DSET}} &\cong 
    \sum_\ell H_{\mathrm{SN}}^{(\ell)}
    - \Delta \sum_e \frac{1}{|\widehat{G}|}\sum_{\chi \in \widehat{G}} \chi(g)  U_e(\chi)
    - \Delta' \sum_{c} \frac{1}{|G|} \sum_{g \in G} \prod_{p \in c} (B_p^{g})^{\pm 1}
    \\
    &= H_{\mathrm{CN}} ,
\end{align}
\end{widetext}
where the $(\pm 1)$ superscript is determined by the relative orientation of $p$, as in Eq.~\eqref{eq:SMNCube}. Here, we have used Eq.~\eqref{eq:DisentanglingHGSN} and 
\begin{align}
    U_{\mathrm{3DSET}}^\dagger \,F_c\, U_{\mathrm{3DSET}}
    =
    \frac{1}{|G|} \sum_{g\in G}\prod_{p\in c} L_p^{\pm 1}(g)\, (B_p^{g})^{\pm 1} .
\end{align}
Again, the gapped-groundspace-preserving equivalence relation allows changes in the choice of local terms and removal of ancilla in (many-body) product states. 
In particular, we have projected out all plaquette degrees of freedom in the $\ket{+}$ state. 
Above, we have $U_e(\chi)=\widehat{\chi}_{e,{\ell}} \widehat{\chi}_{e,{\ell'}}$. In this case, we choose not to project out energetically forbidden local degrees of freedom, corresponding to pairs of string degrees of freedom in $\mathcal{C}_g,\mathcal{C}_{g'}$, with $g g' \neq 1$, on layers that intersect a common edge. 
Instead, we leave an edge term in the cage-net Hamiltonian that energetically penalizes strings with unmatched sectors on an edge. 
Here, $H_{\mathrm{SN}}^{(\ell)}$ denotes the string-net Hamiltonian for $\mathcal{C}_1$, acting in the larger edge Hilbert space corresponding to $\mathcal{C}_G$, including vertex terms that energetically enforce the fusion rules of $\mathcal{C}_G$ strings. 

It is known that pairs of $\mathcal{C}_g,\mathcal{C}_{g'}$ strings on a common edge with $g g' = 1$ correspond to lineon operators. 
In this sense, a single string corresponds to half a lineon operator. 
From this, we can see that the extended line-like excitations described above are composite excitations that can be decomposed into short segments of half-lineon strings. 
The full excitation spectrum of the gauged cage-net model descends from the emergent anyons of the string-net layers, as described in Sec.~\ref{sec:general}.

\paragraph*{Cage-net ground space degeneracy via gauging:} The gauging formulation of general Abelian $p$-string condensation provides a method to compute the ground space degeneracy of all cage-net models. 
The ground space degeneracy of the gauged model is given by counting the dimension of the symmetric subspace within each inequivalent symmetry-twisted ground state sector of the original model. 
Here, the symmetry condition requires invariance under all local and global 1-form $\widehat{G}$ symmetry operators. 
In this setting, equivalence classes of symmetry-twisted ground spaces are described by the first homology group of the manifold $H_1(\widehat{C},\widehat{G})$ or equivalently, the second cohomology group $H^2(C,\widehat{G})$, corresponding to inequivalent, homologically nontrivial $p$-string configurations. 
Each layer that such a homologically nontrivial $p$-string passes through lies in a symmetry-twisted sector corresponding to ground states of the relevant string-net in the presence of a $\widehat{G}$ boson. 
The ground space degeneracy of the string-membrane-net, or equivalently the cage-net, model can then be written as
\begin{align}
        \mathrm{Tr} (\Pi^{0}_{\mathrm{CN}}) = \sum_{[c] \in H^2(C,\widehat{G})} \mathrm{Tr} \Big( \Pi_{\widehat{G}^{(1)}} \prod_{\ell} (\Pi^{c\,\cap\,  \ell}_{\mathrm{SN}^{(\ell)}} ) \Big) ,
\end{align}
where $\Pi_{\widehat{G}^{(1)}}$ denotes the projector onto the symmetric subspace under the diagonal $\widehat{G}$-valued 1-form symmetry, $c\cap \ell$ denotes the $\widehat{G}$ charge induced by $c$ on layer $\ell$, and $\Pi^{c\, \cap \, \ell}_{\mathrm{SN}^{(\ell)}}$ denotes the $c\cap \ell$-twisted string-net ground state projector on layer $\ell$.

For a three dimensional torus, i.e.~periodic boundary conditions along each axis, the symmetry projector $\Pi_{\widehat{G}^{(1)}}$ can be written as a product of local and global symmetry projectors. 
The string-net ground space projectors automatically satisfy the local symmetry projectors, but only a subspace of the ground states satisfy the global symmetry projectors. 
The global symmetry projector in 3-dimensional space can be expanded into a product of symmetry projectors for a basis of cohomologically nontrivial 1-cocycle representatives.
The symmetry projector for a generating 1-cocycle can be expanded as a product of global symmetry-sector projectors on individual layers, that satisfy a constraint that the product of all sector labels must be trivial. 
For a 1-cocycle representative $z\in Z^1(C,\widehat{G})$ of a nontrivial 1-cohomology class, the global symmetry projector $\Pi_{z}$ can be expanded as
\begin{align}
    \Pi_{z} = \sum_{\{ g_\ell \in G | \prod_\ell g_\ell = 1\}}\prod_{\ell} ( \Pi_{z\cap \ell}^{g_\ell} ) ,
\end{align}
where the product is taken over layers $\ell$ that intersect the support of $z$, and $\Pi_{z\cap \ell}^{g_\ell}$ is the projector onto eigenvalue $g_\ell$ of the global symmetry on the restricted 1-cocycle $z\cap \ell \in Z^1(C^{(\ell)},\widehat{G})$ on layer $\ell$. 
The 1-form symmetry projector can then be written as a product of a local symmetry projector and these global projectors for a set $z$ of 1-cocycle representatives for 1-cohomology classes that generate the full first cohomology group
\begin{align}
     \Pi_{\widehat{G}^{(1)}} = \Pi^{\mathrm{loc}}_{\widehat{G}^{(1)}} \prod_z (\Pi_{z} ) .
\end{align}
Finally, the ground space degeneracy of the cage-net model can be written in terms of sums and products of charge and symmetry projectors on the individual string-net layers, that satisfy a global constraint
\begin{align}
    \mathrm{Tr} (\Pi^{0}_{\mathrm{CN}}) = &\sum_{[c] \in H^2(C,\widehat{G})} 
    \sum_{\{ g^z_\ell \in G | \prod_\ell g^z_\ell = 1\}}
    \nonumber \\
    &\qquad \prod_{\ell} \mathrm{Tr} \Big( \Pi^{c\,\cap\,  \ell}_{\mathrm{SN}^{(\ell)}}\prod_{z \cap \ell} (\Pi_{z\cap \ell}^{g^z_\ell} )  \Big) ,
\end{align}
where $z\in Z^1(C\widehat{G})$ form a minimal generating set of cohomologically nontrivial 1-cocycles. 
The projector $\Pi^{c\,\cap\,  \ell}_{\mathrm{SN}^{(\ell)}}\Pi_{z\cap \ell}^{g_\ell}$ picks out the string-net ground states, in the symmetry-twisted sector that corresponds to a pinned $c\cap \ell\in \widehat{G}$ boson, where the total $G$-sector of all strings passing through the $z$ 1-cocycle is $g_\ell \in G$. 

We comment that it is also possible to construct hybrid models with a mixture of fully mobile and restricted mobility particles by performing $p$-string condensation on subgroups. An example would include the lineonic hybrid X-cube model, studied on the lattice in Ref.~\onlinecite{hybrid2021} and as a field theory in Ref.~\onlinecite{HsinSlagle21}.

\subsection{Generalizations}

The 1-form gauging construction of string-membrane-net and cage-net models we have described above can be extended by viewing it in terms of a more general coupling procedure between 2+1D topological layers and a 3+1D topological bulk. This more general construction requires an Abelian 1-form symmetry in a 3+1D topological bulk state which is then stacked with layers of 2+1D topological orders sharing the same 1-form symmetry group; $p$-string excitations created by truncated diagonal symmetry operators across the bulk and layers are then condensed. Alternatively, this condensation can again be induced via gauging the composite 1-form symmetry which allows for the construction of string-membrane-net models with non-Abelian fractons, inherited from non-Abelian point particles in the 3+1D bulk. Appropriate 3+1D bulk models can be found on the lattice using either untwisted Dijkgraaf-Witten gauge theories for potentially non-Abelian groups with central Abelian subgroups that lead to a 1-form symmetry. The point-particles with string-operators that are charged under the 1-form symmetry become fractons after the $p$-string condensation. Similarly, Walker-Wang lattice models~\cite{walker2012} based on $\mathrm{Rep}(G)$ can be used, where an Abelian grading leads to a 1-form symmetry. The string types in the nontrivial graded sectors then become fractons after $p$-string condensation. 

The coupling of a 3+1D bulk topological order to 2+1D Abelian topological layers can also be implemented via gauging planar subsystem symmetries of the 3+1D bulk, as discussed in Ref.~\cite{designer}. Furthermore, the construction we have outlined above with non-Abelian layers can be viewed as gauging a diagonal planar subsystem symmetry on an appropriate 3+1D bulk that has been stacked with layers of 0-form symmetry enriched topological orders. The coupling picture can then be used to derive topological defect network representations~\cite{defectnetworks} of generalized string-membrane-net models. 

All the $p$-string condensations described above are Abelian, even in cases where the resulting fracton theory is non-Abelian. 
It is possible, in principle, to condense non-Abelian $p$-strings. This was attempted in Ref.~\cite{nonabelian}: however, the resulting model did not lead to a consistent $p$-string condensation as the $p$-strings were not fluctuated in a coherent way. Unlike Abelian $p$-strings, it is more challenging to coherently fluctuate non-Abelian $p$-strings which require nontrivial morphisms to describe their fusion and splitting vertices. A potential approach to implement coherent fluctuations of non-Abelian $p$-strings is as follows: consider a stack of 2+1D string-net models, with $p$-strings formed by an algebra of mutual bosons that is supported fully on the plaquettes of the string-nets. For example, this includes the algebra generated by $1, \tau \bar{\tau}$ in the doubled Fibonacci topological order, $1,\psi\bar{\psi},\sigma\bar{\sigma}$ in the doubled Ising topological order, and the charges $1,[-],[2]$, in the $\mathrm{Rep}(S_3)$ gauge theory. Next, we introduce a background 3+1D model that the $p$-string excitations can be coupled to such that they can be coherently fluctuated. A suitable choice is the Walker-Wang model based on the emergent anyon theory of a single string-net layer. The Walker-Wang model can then be driven into a superposition state corresponding to the $p$-strings by energetically enforcing the fusion morphism of the appropriate algebra at each vertex. Finally, terms that couple the $p$-string-net configurations in the Walker-Wang model to matching plaquette excitations on the layers are added, along with hopping terms that fluctuate the $p$-string excitations on the 2+1D layers. In the strong coupling limit of the above terms, the Walker-Wang model should be reduced to a coherent sum over $p$-string configurations, which are strongly coupled to the layers and create physical $p$-string excitations matching the configurations in the Walker-Wang model. 

The above $p$-string condensation procedure can be simply implemented in a tensor network model. This can be achieved by projecting the Walker-Wang ground state~\cite{walker2012} onto the fusion morphisms of the appropriate condensing algebra at each vertex~\cite{Kong2014}, and then replacing the plaquette tensors in the string-net ground state tensor network~\cite{Gu2009,Buerschaper2009} with controlled plaquette tensors that introduce appropriate anyonic excitations~\cite{Bultinck2015,Williamson2017}. These controlled plaquette tensors should be chosen to introduce a plaquette excitation that matches the state of the edge in the projected Walker-Wang state that passes through the plaquette. While the above procedure suggests coherent non-Abelian $p$-string condensation is possible in principle, it remains a challenge to analyse the superselection sectors of the resulting fracton order. The underlying reason for this is that in more general non-Abelian condensations, there can be fusion channels of seemingly nontrivial excitations into the condensate. There can also be related splitting of anyons into combinations of trivial and nontrivial defects sectors. For examples such as doubled Fibonacci layers, it is hence unclear if any nontrivial superselection sectors remain after $p$-string condensation. This suggests generalizing $p$-string condensation to non-Abelian groups such as $\mathrm{Rep}(S_3)$ is a more modest goal, as $\mathrm{Rep}(G)$ condensation does not lead to the splitting of anyons into distinct defect sectors. We leave a thorough investigation of the algebraic theory of non-Abelian $p$-string condensation, and the resulting fracton orders, to future work.


\section{Discussion}
\label{sec:cncls}

In this work, we have shown how gauging an Abelian 1-form symmetry on a stack of 2+1D topological orders embedded in 3+1D implements $p$-string condensation, thereby providing a new algebraic perspective on fracton phases that, in contrast to the original formulation of $p$-string condensation, does not require traversing a phase transition. In particular, here we have shown how this procedure can be carried out directly on the lattice and in field theory. For the latter, we have discussed two concrete examples, the 2+1D $\mbZ_N$ Plaquette Ising model and the 3+1D $\mbZ_N$ X-Cube model. In both cases, we gauge a diagonal subgroup of a foliated symmetry, and interpret the resulting phase as condensing $p$-string operators. We have used the foliated field theory perspective in studying these examples, such that there is still a non-zero separation between the underlying layers: strictly speaking, the resulting gauged theory is not a truly continuum theory. It would be interesting to understand the continuum limit $\varepsilon_i \rightarrow 0$ of these theories when the layers are brought infinitesimally close together. For the particular case when the theories on the layers and in the bulk are all Abelian gauge theories, this question was partially addressed in Ref.~\cite{Slagle21}, but the continuum limit with general TQFTs on the layers remains an open question and is beyond the scope of this paper. On the lattice, we have shown how this gauging procedure can be carried out more generally by considering string-net models on the 2+1D layers and gauging a 1-form symmetry in 3+1D that is generated by Abelian anyons of the underlying 2+1D TQFTs. This formulation of $p$-string condensation provides an exact gap-preserving procedure for producing fracton phases from stacks of lower dimensional topological orders and, as we discuss in a forthcoming work~\cite{forthcoming}, uncovers a web of dualities between various topological and fractonic phases of matter. 

On the lattice, gauging generalized Abelian symmetries can be implemented via adaptive finite-depth local unitary (AFDLU) circuits~\cite{Williamson2020a,Tantivasadakarn2021}. Recently, coarse classifications of topological phases have been proposed that allow equivalence up to AFDLU~\cite{Tantivasadakarn2022}. Our results demonstrate that all cage-net models are in the same AFDLU equivalent phase as decoupled layers of string-net models. If the string-net layers host solvable anyons, existing results show that the resulting cage-net is AFDLU equivalent to the trivial phase~\cite{Ren2024}. To the best of our knowledge, all 3+1D fracton models that have appeared in the literature are AFDLU equivalent to a stack of decoupled 3+1D blocks, and 2+1D layers of conventional TQFT models (some of the twisted models of Ref.~\cite{twisted} may be an exception but this remains to be seen). It is an interesting open question to construct a fracton model that is not AFDLU equivalent to decoupled layers of TQFT models. We anticipate that such a construction is required to find a non-Abelian type-II model (that is, a model with only immobile non-Abelian topological excitations).  

Another interesting question is to find a classification of all fracton models with a layered structure, such as those constructed here. Even for fracton models where all particles are planons with the same relative orientation, this remains an open question. The classification of such models is particularly relevant as it is equivalent to a classification of spacetime topological phases of Floquet codes that switch between 3+1D TQFT codes~\cite{domkitp}. Finally, it would be interesting to study whether ``classically" gauging~\cite{Ellison2025} the diagonal higher-form symmetries studied in this work can generate interesting mixed-state fracton orders, generalizing recent work on intrinsically mixed-state topological order~\cite{Ellison2025,Sohal2025,hsin2025}.


\stoptoc
\begin{acknowledgements}
We are especially thankful to Nati Seiberg for stimulating discussions and for his enthusiastic encouragement. A.P.~is grateful to Huan He for inspiring discussions at the beginning stages of this project. N.T. is grateful to Xie Chen and Mike Hermele for helpful discussions.  D.W.~acknowledges useful discussions with David Aasen and Kevin Slagle in the early stages of this project. This research was supported in part by grant NSF PHY-2309135 to the Kavli Institute for Theoretical Physics (KITP). P.G.~was supported by the Physics Department of Princeton University and the Simons Collaboration on Global Categorical Symmetries. This material is based upon work supported by the Sivian Fund and the Paul Dirac Fund at the Institute for Advanced Study and the U.S. Department of Energy, Office of Science, Office of High Energy Physics under Award Number DE-SC0009988 (A.P.). N.T. is supported by the Walter Burke Institute for Theoretical
Physics at Caltech. D.W.~is supported by the Australian Research Council
Discovery Early Career Research Award (DE220100625). The authors of this paper were ordered alphabetically.
\end{acknowledgements}


\appendix
\resumetoc

\section{String-net models and their 1-form symmetries}
\label{app:stringnet}

In this Appendix, we review string-net models and their symmetries. In particular, we discuss $G$-graded string-net models and review how gauging a 1-form symmetry generated by Wilson string operators for an Abelian anyon produces a phase that is equivalent (as a topological order) to one where that Abelian anyon has been condensed. 

\paragraph*{Lattice Hamiltonian:} We consider string-net models~\cite{Levin2005} defined on a directed square lattice with string degrees of freedom living on the edges, whose basis is labelled by simple objects $s$ from a unitary fusion category $\mathcal{C}$. The vertex degrees of freedom are given by 
\begin{align}
\bigoplus_{i,j,k,l\in \mathcal{C}} \hom(i\otimes j,k \otimes l),
\end{align} 
which span the fusion and splitting space of the surrounding strings $i,j,k,l$. 
For simplicity, we present the multiplicity free case where the vertex degree of freedom can be identified with an edge that resolves the lattice to be trivalent: 
\begin{align*}
\vcenter{\hbox{\includegraphics[page=5]{TikzFigures}}} \qquad \rightarrow  \qquad \vcenter{\hbox{\includegraphics[page=6]{TikzFigures}}} 
\end{align*}

The string-net Hamiltonian takes the form 
\begin{align}
\label{eq:SNH}
    H_{\mathrm{SN}} = - \sum_v \left(A_v^{(1)}+A_v^{(2)}\right) - \sum_p \sum_{s\in\mathcal{C}} \frac{d_s}{\mathcal{D}^2} B_p^s ,
\end{align}
where the first sum is over all vertices of the square lattice. Here, $\mathcal{D}^2=\sum_{s\in\mathcal{C}} d_s^2$ is the square of the total quantum dimension of $\mathcal{C}$, and $d_s$ is the quantum dimension of $s\in\mathcal{C}$. 
The vertex terms in the Hamiltonian are
\begin{align}
\label{eq:snv}
A_v^{(1)}\ \vcenter{\hbox{\includegraphics[page=1]{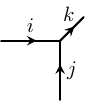}}} &= N_{ij}^{k}\  \vcenter{\hbox{\includegraphics[page=1]{TikzFigures}}} \, , 
\\
A_v^{(2)}\ \vcenter{\hbox{\includegraphics[page=2]{TikzFigures}}} &= N_{i^*j^*}^{k^*}\  \vcenter{\hbox{\includegraphics[page=2]{TikzFigures}}} \, ,
\end{align}
where $N_{i,j}^{k}=\dim( \hom(i \otimes j, k))$ is the dimension of the fusion space which we have assumed to be either 0 or 1, due to the multiplicity free condition. The inverse of a simple object $s$ is defined to be the object $s^*$ such that $N_{s,s^*}^{1}=1$, where the superscript $1$ denotes the vacuum object which satisfies $N_{s,1}^s=N_{1,s}^s=1$. Reversing the direction of an edge in the lattice is implemented by interchanging the basis state for each simple object $s$ with that of its inverse $s^*$. From this, we see that $A_v^{(2)}$ implements a similar constraint to $A_v^{(1)}$ when the surrounding edge orientations are reversed. The vertex terms in the Hamiltonian enforce that only string-net configurations that satisfy the fusion rules at every vertex can appear in the ground space. 

The $B_p^s$ plaquette term is defined within the simultaneous $+1$ eigenspace of the vertex terms by fusing a string of type $s$ into the boundary of the plaquette $p$:
\begin{align}
\label{eq:snp}
B_p^s\ \vcenter{\hbox{\includegraphics[page=4]{TikzFigures}}}  =\ \vcenter{\hbox{\includegraphics[page=3]{TikzFigures}}} ,
\end{align}
while outside the $+1$ subspace of the vertex terms, $B_p^s$ is defined to act trivially as multiplication with 0. We remark that to extract matrix elements from the above definition of $B_p^s$, one must use the fusion rules of the category $C$ to resolve the $a$ string into the boundary of the plaquette~\cite{Levin2005}. To ensure that $B_p^s$ acts nontrivially, the plaquette can be thought of as being punctured. This means that the $s$ string cannot simply be contracted away within the plaquette, but must be fused into the boundary to return to the original lattice. 

\paragraph*{Anyon excitations:} The excitations of a string-net model can be organized into point-like superselection sectors, corresponding to local excitations that are equivalent up to local operators. 
The superselection sectors for a string-net model based on the category $\mathcal{C}$ are given by the simple objects of the Drinfeld center $\mathcal{Z}(\mathcal{C})$. These point-like superselection sectors can be created and moved via string operators, and are formally described by an anyon theory associated to the modular tensor category $\mathcal{Z}(\mathcal{C})$. 
For an input category $\mathcal{C}$ equipped with a modular braiding, the string operators can be represented by strings from $\mathcal{C}$ above and below the lattice. 
These strings define string operators via resolution into the lattice using the modular braiding and fusion of $\mathcal{C}$.
In this case, the emergent anyon theory can be decomposed into a chiral copy of $\mathcal{C}$ represented by strings above the lattice, and an antichiral copy $\mathcal{C}^{\mathrm{rev}}$ represented by strings below the lattice, hence $\mathcal{Z}(\mathcal{C})\cong \mathcal{C} \boxtimes \mathcal{C}^{\mathrm{rev}}$~\cite{Levin2005,koenig2010quantum}. 
More generally, when $\mathcal{C}$ is not modular, it is necessary to solve for a set of half-braidings to construct string operators corresponding to the anyons in $\mathcal{Z}(\mathcal{C})$~\cite{burnell2021gen,Green2023}. 

Each string-net comes equipped with a special subset of anyons that form a maximal set of condensible bosons, known as the canonical Lagrangian algebra object $\mathcal{Z}(1)\in \mathcal{Z}(\mathcal{C})$. These are the anyons that condense at the smooth boundary constructed by treating the input category to the string-net as a module over itself~\cite{kitaevkong}.
These anyons can be identified with irreducible representations of the algebra generated by plaquette operators $B_p^s$. 
As such, they correspond to plaquette-only excitations (also referred to as the pure fluxes of the theory). More generally, an extra dangling string degree of freedom is required to support general anyons from $\mathcal{Z}(\mathcal{C})$ on a plaquette~\cite{dyonicspectrum}. 

We consider the following perturbation of the string-net Hamiltonian 
\begin{align}
    H_\lambda = (1-\lambda) H_{\mathrm{SN}} + \lambda V ,
\end{align}
where $V=-\sum_e \ket{1}\bra{1}_e$ and $\lambda \in [0,1]$. As $\lambda$ increases, the ground space of this Hamiltonian undergoes a phase transition to the trivial phase. This phase transition is driven by the proliferation of plaquette-only excitations, which is described by the condensation of the canonical Lagrangian algebra object~\cite{Bais2009}.

\paragraph*{1-form symmetry:} The Abelian anyons in $\mathcal{Z}(\mathcal{C})$ form a group under fusion, which we denote by $A$. 
The string operators for the Abelian anyons $A$ generate a 1-form symmetry of the string-net Hamiltonian for the input category $\mathcal{C}$. 
This 1-form symmetry is anomalous whenever the group of Abelian anyons $A$ has nontrivial braiding. 
Any subgroup of Abelian anyons $\widehat{G}\leq A$ defines a $G$-grading on the full anyon theory $\mathcal{Z}(\mathcal{C})$. 
This is because the braiding of the Abelian anyons $\chi \in \widehat{G}$ with other anyons in $\mathcal{Z}(\mathcal{C})$ defines a $G$-character
\begin{align}
    M_{a,\chi} = \chi(g) ,
\end{align}
for some $g\in G$~\cite{barkeshli2014symmetry}. 
In this case, $a$ is assigned to the component labelled by $g\in G$, which is denoted by $a_g$. In particular, any Abelian subgroup of a Lagrangian algebra object $\mathcal{Z}(1)$ is inherited from a grading of the input category $\mathcal{C}$~\cite{levin2016set,cheng2017set,Williamson2017}. 
Such a grading is defined by operators
\begin{align}
    \label{eq:chiop}
    \widehat{\chi}_e \ket{s_g}_e = \chi(g) \ket{s_g}_e ,
\end{align}
that act on each edge $e$ of the string-net.

A $G$-grading of $\mathcal{C}$ corresponds to a decomposition 
\begin{align}
    \mathcal{C}=\bigoplus_{g\in G} \mathcal{C}_g , 
\end{align}
where the objects $r\in \mathcal{C}_g, s\in \mathcal{C}_h$ satisfy $r \times s \in \mathcal{C}_{gh}$. 
This defines graded plaquette operators
\begin{align}
    B_p^g = \sum_{s \in \mathcal{C}_g} \frac{d_s}{\mathcal{D}_1^2} B_p^{s} ,
\end{align}
where $\mathcal{D}_1$ is the total quantum dimension of $\mathcal{C}_1$. 
The graded plaquette operators form a representation of $G$ since $B_p^g B_p^h = B_p^{gh}$.
Hence the $B_p^g$ operators can be simultaneously diagonalized with eigenvalues given by characters $\chi \in \widehat{G}$. 
The $\chi$ eigenspace of the $B_p^g$ operators corresponds to the $\chi$ Abelian anyon in $\mathcal{Z}(1)$. 
The $\widehat{\chi}_e$ operators create these Abelian anyons in pairs on the plaquettes adjacent to $e$, since
\begin{align}
    \widehat{\chi}_e B_p^g = \chi^{\pm 1}(g) B_p^g \widehat{\chi}_e
\end{align}
for $e\in p$, where $\pm 1$ reflects whether the orientation of $e$ matches that of $\partial p$. 
These operators define an on-site representation of the anomaly-free $\widehat{G}$ 1-form symmetry 
\begin{align}
    U:\quad Z^1(C,\widehat{G}) &\rightarrow \mathcal{U}(\mathcal{H}) \nonumber
    \\
    z &\mapsto U(z) := \prod_{e} U_e({z_e})
    ,
\end{align}
where $\mathcal{U}(\mathcal{H})$ denotes the group of unitary operators on the edge qudit Hilbert space, and $z_e \in \widehat{G}$ is the coefficient assigned to edge $e\in C$ by the 1-cocycle $z\in Z^1(C,\widehat{G})$. 
Here, the on-site representation of $\chi\in\widehat{G}$ is ${U_e(\chi)=\widehat{\chi}}_e$ (see Eq.~\eqref{eq:chiop}). We remark that this is indeed a 1-form symmetry since $C$ is two-dimensional and hence, by Poincar\'e duality, the group of 1-cocycles is isomorphic to the group of 1-cycles on the cellulation that is dual to $C$. In other words, the 1-form symmetries live on the dual lattice. 

The perturbed Hamiltonian 
\begin{align}
    H_\lambda = (1-\lambda) H_{\mathrm{SN}} + \lambda V_{\widehat{G}} ,
\end{align}
where $V_{\widehat{G}}=-\sum_e \frac{1}{|\widehat{G}|}\sum_{\chi \in \widehat{G}}U_e(\chi)$ and $\lambda\in [0,1]$, undergoes a $\widehat{G}$-boson condensation phase transition to the string-net model described by the subcategory $\mathcal{C}_1$ as $\lambda$ is tuned from 0 to 1. 

\paragraph*{Gauging the 1-form symmetry:} The $\widehat{G}$-boson condensation described above can be implemented directly via gauging the 1-form $\widehat{G}$ symmetry; this is because gauging a symmetry condenses the defects that are created at the boundary of truncated symmetry operators. Here, by definition, truncated 1-form $\widehat{G}$ symmetry operators create $\widehat{G}$-bosons at their boundary and so gauging the symmetry condenses these bosons. Note that unlike the anyon condensation procedure discussed above, this gauging procedure does not involve any phase transition and is an exact gap-preserving map between the un-condensed and condensed phases.

To gauge the 1-form symmetry, we first introduce $\mathbb{C}[\widehat{G}]$ degrees of freedom in the state $\ket{+}$ onto the plaquettes $p \in C$, which correspond to vertices in the dual cellulation. 
Here, the $\ket{+}$ state corresponds to the trivial character in $\widehat{G}$. 
The plaquette degrees of freedom play the role of generalized gauge fields for the 1-form symmetry. 
Next, we measure the set of projection operators
\begin{align}
    \label{eq:GGauss}
    \Pi_e(g) := \frac{1}{|\widehat{G}|}\sum_{\chi \in \widehat{G}} \chi^*(g) U^\dagger_p(\chi) U_e(\chi) U_{p'}(\chi) ,
\end{align}
for $g\in G$, on every edge $e$, and for the adjacent plaquettes $p,p'$ in $C$. We assign the same right handed orientation to all plaquettes, such that $p$ is the plaquette with orientation matching $e$ and $p'$ is the plaquette with orientation opposite to $e$. 
The plaquette operators are defined as follows: $U_p(\chi) \ket{g} = \chi(g) \ket{g}$, where the edge operators are defined above. 
The above measurements detect violations of the generalized Gauss's law that attaches edge 1-form symmetry charges to generalized field lines on the adjacent plaquettes. 
This follows from the action of the $\Pi_e(g)$ operators in the basis of $G$ group elements on plaquettes $p,p'$, and $\mathcal{C}$ string types on edge $e$,
\begin{align}
    \Pi_e(f) \ket{g_p\, (s_h)_e\, k_{p'}} = \delta_{f^{-1} g^{-1} h k}\ket{g_p\, (s_h)_e\, k_{p'}} ,
\end{align}
where $f,g,h,k\in G$ and $s_h\in \mathcal{C}_h$. 

The standard 1-form gauging procedure corresponds to applying the projector $\Pi_e(1)$ to all edges, which can be achieved via post-selection. 
Alternatively, this can be achieved efficiently by applying an Abelian byproduct operator that depends on the measured outcomes of the projection operators in Eq.~\eqref{eq:GGauss}. We now explain how to find the byproduct operator. 

On the subspace of symmetric states, the outcomes $g_e$ of the measurement described in Eq.~\eqref{eq:GGauss} must form a $G$-valued 1-cycle on $C$. 
This is because the product of projectors on edges $e_1,e_2,e_3,$ adjacent to an arbitrary trivalent vertex $v$ satisfy
\begin{align}
    \Pi_{e_1}(g_{1})\Pi_{e_2}(g_{2})\Pi_{e_3}(g_{3}) 
    U_{e_1}(\chi) U_{e_2}(\chi) U_{e_3}(\chi^*) \nonumber \\
    =
    \chi(g_1 g_2 g_3^{-1})
    \Pi_{e_1}(g_{1})\Pi_{e_2}(g_{2})\Pi_{e_3}(g_{3}) .
\end{align}
Here, we have taken $e_1,e_2,$ to be oriented towards $v$ and $e_3$ to be oriented away from $v$, with other configurations following analogously. 
The above equation follows from the fact that each projector satisfies 
\begin{align}
    \Pi_e(g) U_e(\chi) = \chi(g) U_p(\chi) U^\dagger_{p'}(\chi) \Pi_e(g) .
\end{align} 
Hence, for any state that satisfies the 1-form symmetry $U_{e_1}(\chi) U_{e_2}(\chi) U_{e_3}(\chi^*)$, the measurements of $\Pi_{e_1},\Pi_{e_2},\Pi_{e_3}$ must result in outcomes that satisfy $g_1 g_2 g_3^{-1}=1$ for the post-measurement state to be nonzero. 

To simplify the description of the byproduct operator, we consider a situation with trivial $G$-valued 1st homology, such as the infinite plane. 
The byproduct operator we describe also applies to states that satisfy all local and global 1-form symmetries on cellulations with nontrivial 1st homology. 
By assumption, the outcomes of gauging measurements on all edges $g_e$ that form a $G$-valued 1-cycle on $C$ must in fact form a 1-boundary. 
Hence, there is a 2-chain $h_p$ on $C$ that satisfies $(\partial h_p)_e = g_e$. 
The byproduct operator can be written as $\prod_p L_p (h_p)$. 
The application of this operator after gauging has the sole effect of setting the values of all edge projectors to $1$. 
This follows from the commutation relations 
\begin{align}
    L_p(g) \Pi_e(g) &= \Pi_e(1) L_p(g) , \\
    L_{p'}^\dagger(g) \Pi_e(g) &= \Pi_e(1) L_{p'}^\dagger(g) ,
\end{align} 
where $p/p',$ are adjacent to $e$ with matching/mismatching orientation, respectively. 
We also use the symmetry of the initial plaquette states $L_g \ket{+} = \ket{+}$, which can be expanded in the $G$-basis as
\begin{align}
    \ket{+}=\frac{1}{|G|} \sum_{g\in G} \ket{g} .
\end{align}
The choice of 2-chain $h_p$ is arbitrary, even in the situation with trivial 2nd homology, where there are multiple inequivalent choices. This is due to the symmetry assumption on the input states. 

We now describe the gauging of operators: without loss of generality, we consider an operator $M$ that has a definite 1-form charge on each edge described by the 1-chain $g \in Z_1(C,G)$. 
Any operator can be written as a sum of such operators, and the gauging procedure can be applied to each individually, as it is a linear map. 
If the operator $M$ has support on a subregion $R\subset C$, the cochain $g$ necessarily takes the value $g_e=1$ on all edges $e\notin R$. 
If the operator $M$ is symmetric, then the 1-chain $g$ must be a 1-cycle. 
If $M$ is both symmetric and local, then $M$ must be a 1-boundary (as there are no nontrivial local 1-cycles). 
In this case, we can find a 2-chain $h \in Z_2(C,G)$ that satisfies $\partial h = g$. 
There is freedom when choosing $h$ as it can be multiplied by an arbitrary 2-cycle. 
Here, we pick the unique $h$ that has local support, within a neighborhood of $R$. 
Now, for a local symmetric operator $M$, the gauged operator is
\begin{align}
\label{eq:GuugingM}
    \mathcal{G}(M) := M \prod_{p\in C} L_p(h_p) ,
\end{align}
where $L_p(h_p)$ denotes left multiplication by $h_p$ on $p$, and $h_p$ is trivial outside a neighborhood of $R$. 
Due to the choice of 2-chain, $\mathcal{G}(M)$ is supported on a neighborhood around $M$. 
Hence, the gauging process maps local symmetric operators to local symmetric operators. 
In the case of an operator with trivial 1-form charge on all edges, the gauged operator is the same as the original operator.

Gauging the 1-form $\widehat{G}$ symmetry of the string-net Hamiltonian $H_{\mathrm{SN}}$ is performed term-wise on the vertex and plaquette operators. 
More precisely, we rewrite the plaquette operators as sums of $B_p^g$ operators
\begin{align}
    \sum_{s\in\mathcal{C}} \frac{d_s}{\mathcal{D}^2} B_p^s = \frac{1}{|G|}\sum_{g \in G} B_p^g .
\end{align}
This leads to 
\begin{align}
\label{eq:GSN}
    H_{\mathrm{GSN}} =& - \sum_v \mathcal{G}(A_v^{(1)})+\mathcal{G}(A_v^{(2)}) - \sum_p \frac{1}{|G|}\sum_{g\in G}  \mathcal{G}(B_p^g) 
    \nonumber \\
    &- \Delta \sum_e \Pi_e(1), 
\end{align}
where the $\Pi_e(1)$ terms are projectors onto states satisfying the generalized Gauss's law (see Eq.~\eqref{eq:GGauss}) and $\Delta>0$ controls the energy scale of states that violate the Gauss's law. 
The vertex terms are unchanged, $\mathcal{G}(A_v)=A_v$, since $A_v$ is diagonal in the on-site symmetry basis. 
The plaquette operators satisfy $\mathcal{G}(B_p^s)= L_p(g) B_p^s$, for $s\in\mathcal{C}_g$, and hence $\mathcal{G}(B_p^g)= L_p(g) B_p^g$, by linearity of $\mathcal{G}$.
Finally, we can write the 1-form gauged string-net Hamiltonian 
\begin{align}
\label{eq:GSNH}
    H_{\mathrm{GSN}} =& - \sum_v A_v^{(1)}+A_v^{(2)} - \sum_p \frac{1}{|G|} \sum_{g \in G}   L_p(g) B_p^g
    \nonumber \\
    &- \Delta \sum_e \Pi_e(1). 
\end{align}
This model is equivalent to a $G$-symmetry-enriched string-net model (see Refs.~\cite{levin2016set,cheng2017set}). The global $G$ symmetry of this model is represented by operators $\prod_p L_p(g)$\footnote{\label{ftnt:app}Here, we are using the fact that $G$ is an Abelian group. For non-Abelian groups, with the choice of conventions we have made, the global symmetry action corresponds to right multiplication.}. The emergent symmetry-enriched theory is described by the relative centre~\cite{Williamson2017,gelaki2009centers}
\begin{align}
\mathcal{Z}_{\mathcal{C}_1}(\mathcal{C})=\bigoplus_{g\in G}\mathcal{Z}_{\mathcal{C}_1}(\mathcal{C}_g) ,
\end{align} 
which was denoted $\mathcal{Z}_{G}(\mathcal{C})$ in Ref.~\cite{Williamson2017}. This emergent symmetry-enriched topological order can be derived from the transformation of the $\mathcal{Z}(\mathcal{C})$ anyons under the $\widehat{G}$ 1-form symmetry as follows:
\begin{itemize}
    \item $\widehat{G}$ bosons corresponding to the 1-form symmetry are condensed and become trivial anyons, equivalent to the vacuum. 
    Hence, any anyons or defects related by fusion with $\widehat{G}$ bosons become equivalent. 
    The condensed $\widehat{G}$ bosons become local charges under the global $G$ symmetry of the symmetry-enriched string-net. 
    \item $\mathcal{Z}(\mathcal{C})$ anyons that have trivial 1-form charge become anyons described by $\mathcal{Z}(\mathcal{C}_1)$. 
    \item $\mathcal{Z}(\mathcal{C})$ anyons that have nontrivial 1-form charge $g\in G$ become $g$-defects in the sector described by $\mathcal{Z}_{\mathcal{C}_1}(\mathcal{C}_g)$. 
\end{itemize}

The ground states of the gauged model can be found by gauging symmetric ground states of the original model in all symmetry-twisted sectors. 
Here, symmetric states correspond to those which are invariant under all $\widehat{G}$ string operators, including the topologically nontrivial strings. 
The symmetry-twisted sectors of the original model correspond to nontrivial homology, or cohomology, classes of symmetry defects. 
These become ground states after gauging, because the symmetry defects are condensed during the gauging process. 
For the $\widehat{G}$ 1-cocycle symmetries considered above, symmetry defects correspond to $\widehat{G}$-bosons.
A basis of inequivalent 2-cocycles is then labelled by configurations with a single plaquette taking a value in $\widehat{G}$, corresponding to states that support a single boson excitation. 
To gauge these states consistently, the state on the plaquette variables must be modified to a corresponding 2-cocycle. 
This is discussed for gauging global symmetries in Appendix~G of Ref.~\cite{williamson2014matrix}. 

The $G$-symmetry-enriched string-net model with input $G$-graded category~$\mathcal{C}_G$ is generalized-local-unitary (gLU) equivalent~\cite{Chen2010} to the standard string-net model with input category~$\mathcal{C}$. 
The SET-entangler circuit that maps the standard string-net to the SET string-net is a product of controlled plaquette $B_p^g$ operators:
\begin{align}
\label{eq:USET}
    U_{\mathrm{SET}} &=\prod_{p} CB_p , \\
    C B_p \ket{g}\ket{\Psi} &= \ket{g} B_p^g \ket{\Psi}.
\end{align}
Here, we view $B_p^g$ as a unitary operator by extending it to act as the identity on states outside of the support space of $B_p^1$. 
Conjugating the $G$-symmetry-enriched string-net Hamiltonian by the disentangling circuit yields 
\begin{align}
    U_{\mathrm{SET}}^\dagger H_{\mathrm{GSN}}U_{\mathrm{SET}} &\cong  H_{\mathrm{SN}} - \sum_{p} \Big( \frac{1}{|G|} \sum_{g\in G} L_p(g) \Big) B_p^1 
    \nonumber \\
    & \quad - \Delta \sum_e \frac{1}{|\widehat{G}|}\sum_{\chi \in \widehat{G}} \chi(g)  U_e(\chi)
    \nonumber \\
    & \cong H_{\mathrm{SN}} - \sum_{p} \Big( \frac{1}{|G|} \sum_{g\in G} L_p(g) \Big) 
    \nonumber \\
    & \cong H_{\mathrm{SN}}
    \label{eq:DisentanglingHGSN}
\end{align}
Here, the equivalence relation captures a change in the choice of local terms that preserves the gapped groundspace, the removal of ancilla qudits that are in a product state, and the projecting out of energetically forbidden local string degrees of freedom in $\mathcal{C}_g$, where $g\neq 1$, on each edge. 

The ground space degeneracy of the gauged model on a torus is given by the number of distinct deconfined anyons after condensation. 
Alternatively, the ground space degeneracy can be calculated by counting the dimension of the symmetric subspace of the original model's ground space in all inequivalent symmetry-twisted sectors. 
This method applies to general manifolds with boundaries. 
Here, the symmetric subspace only includes states that transform trivially under all local, and global, 1-form symmetry operators. 
That is, the symmetric states must be invariant under braiding a $\widehat{G}$ boson around any handle of the manifold.
For a 1-form symmetry, the distinct symmetry-twisted sectors are labelled by $\chi\in\widehat{G}$, with each element determining a ground space in the presence of a single pinned $\widehat{G}$ boson. 
The ground space degeneracy of the gauged model can be written 
\begin{align}
    \mathrm{Tr} (\Pi^{0}_{\mathrm{GSN}}) = \sum_{\chi \in \widehat{G}} \mathrm{Tr} ( \Pi_{\widehat{G}^{(1)}}\Pi^{\chi}_{\mathrm{SN}} ) ,
\end{align}
where $\Pi^{0}_{\mathrm{GSN}}$ is the ground state projector of the gauged string-net model, $\Pi_{\widehat{G}^{(1)}}$ is the projector onto the 1-form symmetric subspace, and $\Pi^{\chi}_{\mathrm{SN}}$ is the projector onto the $\chi$-twisted ground state sector of the string-net. 
On a torus, the global symmetry projector enforces that the overall $G$-sector of the strings in the string-net ground state wave function that pass through either handle of the torus must be trivial. 



\stoptoc
\bibliography{library}


\end{document}